\pdfoutput=1
\documentclass[conference]{IEEEtran}
\IEEEoverridecommandlockouts
\usepackage{cite}
\usepackage{amsmath,amssymb,amsfonts}
\usepackage{algorithm}
\usepackage{algpseudocode}
\usepackage{graphicx}
\usepackage{textcomp}
\usepackage{xcolor}
\usepackage{booktabs}
\usepackage{multirow}
\usepackage{adjustbox}
\usepackage{tikz}
\usetikzlibrary{arrows.meta,positioning,fit,backgrounds,calc,shapes.geometric}
\usepackage[hidelinks]{hyperref}
\graphicspath{{figures/}}

\newcommand{\concat}{\mathbin{\|}}
\newcommand{\Hs}{\mathrm{SHA256}}
\newcommand{\finding}[2]{\par\smallskip\noindent\fbox{\parbox{\dimexpr\columnwidth-2\fboxsep-2\fboxrule}{\small\textit{Finding #1.} #2}}\par\smallskip}
\makeatletter
\newcommand{\linebreakand}{\end{@IEEEauthorhalign}\hfill\mbox{}\par\mbox{}\hfill\begin{@IEEEauthorhalign}}
\makeatother
\algrenewcommand\algorithmicrequire{\textbf{Input:}}
\algrenewcommand\algorithmicensure{\textbf{Output:}}

\begin{document}

\title{Merkle-Tree-Based Verifiable Public Data Auditing for Cloud Object Storage Using Smart Contracts}

\author{\IEEEauthorblockN{Siri Reddy Tadi}
\IEEEauthorblockA{\textit{Dept. of Information Technology} \\
\textit{Chaitanya Bharathi Institute of Technology}\\
Hyderabad, Telangana, India \\
tadi.sirireddy@gmail.com}
\and
\IEEEauthorblockN{K. Gangadhara Rao}
\IEEEauthorblockA{\textit{Dept. of Information Technology} \\
\textit{Chaitanya Bharathi Institute of Technology}\\
Hyderabad, Telangana, India \\
kgangadhar\_it@cbit.ac.in\\
\textit{(Corresponding Author)}}
\linebreakand
\IEEEauthorblockN{M. Venu Gopalachari}
\IEEEauthorblockA{\textit{Dept. of Information Technology} \\
\textit{Chaitanya Bharathi Institute of Technology}\\
Hyderabad, Telangana, India}
\and
\IEEEauthorblockN{C. Chetann Reddy}
\IEEEauthorblockA{\textit{Dept. of Information Technology} \\
\textit{Chaitanya Bharathi Institute of Technology}\\
Hyderabad, Telangana, India \\
chetann0717@gmail.com}
\linebreakand
\IEEEauthorblockN{Jaswanth Bhimireddy}
\IEEEauthorblockA{\textit{Dept. of Information Technology} \\
\textit{Chaitanya Bharathi Institute of Technology}\\
Hyderabad, Telangana, India \\
jaswanthbhimireddy5@gmail.com}
}

\maketitle

\begin{abstract}
Owners of data in cloud object storage cannot directly check whether the provider still keeps every object unchanged. Existing research mostly checks blocks inside files or relies on third-party auditors, and less attention has been given to checking each object against a public blockchain reference while measuring the cost. A full-data audit downloads every object and compares it with a saved hash list, which is slow and only as reliable as that list. We propose a public auditing system in which the SHA-256 hash of each object and its name forms a leaf of a Merkle tree, so the whole dataset is represented by one Merkle root. This root is stored in an Ethereum smart contract as a trusted public record, while the objects stay with the provider. An auditor downloads an object, rebuilds the root using a short Merkle proof and compares it with the stored root, marking the object as intact, modified or missing. In five experiments on an S3-compatible store and a local Hardhat blockchain (normal audit, single- and multiple-object tampering, deletion and scalability), the system detected 15 of 15 single modifications, 2,205 of 2,205 modified objects and 830 of 830 deleted objects with no false alarms. The proof needed only 14 hashes for 10,000 objects, so checking one object took about 10 ms and a 100-object sample about 150 ms regardless of dataset size, while a full audit of 10,000 objects (14.3 s) took about as long as downloading them (13.5 s). Storing the root cost 91,863 gas, 252 times less than storing 1,000 hashes on-chain. The system thus shows exactly which objects were modified or are missing, with its main savings coming from checking selected objects.
\end{abstract}


\section{Introduction}

Organizations increasingly keep unstructured data in cloud object stores whose hardware they neither own nor operate~\cite{armbrust2010view}. Objects can be damaged or lost through disk faults, software bugs, operator mistakes or deliberate interference, and as long as nobody checks, an overwritten or deleted object looks no different to its owner from one that is still intact. An integrity audit therefore needs two ingredients: a description of the correct state that the provider is unable to alter, and a procedure, affordable enough to run regularly, that compares the stored data with that description.

The most direct procedure is a full-data audit, in which the auditor fetches every object and compares its SHA-256 digest with a digest list kept locally. The work grows with the volume of data, failures can be pinpointed only if one digest per object is retained, and the result can be trusted only as far as the stored list can. A Merkle tree~\cite{merkle1980protocols,merkle1987} compresses the digests of $n$ objects into one 32-byte root and allows any single object to be checked against that root with a proof of about $\log_2 n$ hashes. If the root is recorded in a smart contract, it becomes a public reference that neither the provider nor the auditor can silently change, while the objects themselves stay off-chain.

This paper designs, implements and measures such a mechanism for an S3-compatible object store. Its aim is not to propose a new cryptographic primitive, but to establish experimentally what the combination of SHA-256, a Merkle tree and an on-chain root actually provides: which integrity violations it detects, what each audit costs in time and transferred data, how these costs scale with the number of objects, and what the blockchain adds in gas and latency.

\subsection{Research Questions}
\begin{itemize}
\item \textbf{RQ1 (correctness).} Does the audit classify unmodified objects as intact and a single modified object as modified, for different kinds of modification?
\item \textbf{RQ2 (multiple violations).} Does the audit identify every modified and every deleted object when several are affected, distinguish the two cases, and with what probability does a sampled audit of $c$ objects detect $t$ violations?
\item \textbf{RQ3 (scalability).} How do tree construction, proof generation, proof verification and audit time grow with the number of objects $n$?
\item \textbf{RQ4 (communication).} How much data must an auditor retrieve compared with a full-data audit?
\item \textbf{RQ5 (blockchain overhead).} What gas and latency does keeping the trusted root and the audit log on-chain add?
\end{itemize}

\subsection{Contributions}
\begin{enumerate}
\item A Merkle-tree commitment for object storage in which each leaf binds an object's key to its SHA-256 digest, internal and leaf nodes are domain-separated, and unpaired nodes are promoted rather than duplicated (Section~\ref{sec:system}).
\item A smart contract that stores only the root, the object count and an append-only audit log, and that can also verify a Merkle proof on-chain; the off-chain tree is treated as untrusted (Sections~\ref{sec:system}--\ref{sec:impl}).
\item A public audit procedure that downloads the challenged objects themselves, recomputes their leaves and classifies each object as \emph{intact}, \emph{modified} or \emph{missing}, in full or sampled form.
\item A measured evaluation on a local testbed covering normal audits, single- and multiple-object tampering, deletion, scalability from 10 to 10{,}000 stored objects (and to $2^{20}$ leaves in memory), communication volume and smart-contract gas, with five repetitions per configuration (three for the largest dataset) (Sections~\ref{sec:method}--\ref{sec:results}).
\end{enumerate}

\textit{Organization.} Section~\ref{sec:background} gives background and Section~\ref{sec:related} discusses related work. Section~\ref{sec:problem} states the problem, threat model and requirements, Section~\ref{sec:system} presents the system and Section~\ref{sec:impl} its implementation. Section~\ref{sec:method} describes the experimental methodology and Section~\ref{sec:results} the results. Section~\ref{sec:security} analyses security, and Sections~\ref{sec:discussion}--\ref{sec:conclusion} discuss the findings, threats to validity, limitations, future work and conclusions.

\section{Background}\label{sec:background}

This section reviews the building blocks the mechanism relies on. Each subsection ends by stating the role the component plays in our design, so that established material is kept apart from what this paper contributes.

\subsection{Cloud Object Storage and the Trust Problem}
In an object store, every item is an opaque byte sequence that carries some metadata and is addressed by a key in a flat namespace; clients manipulate it through a small set of HTTP requests (PUT, GET, DELETE, LIST). The request vocabulary of Amazon S3~\cite{awss3} has become a common interface that other servers, MinIO among them~\cite{minio}, reimplement. Under the service models described by NIST~\cite{nist800145} and in the early analysis of cloud computing by Armbrust et al.~\cite{armbrust2010view}, the tenant decides what is stored but does not operate the machines that store it. Kamara and Lauter~\cite{kamara2010crypto} therefore argue that a tenant should be able to check its outsourced data cryptographically rather than rely on the provider's word. For integrity, the difficulty is easy to state: a successful GET shows only that \emph{some} bytes came back, not that they are the bytes that were uploaded, and the loss of an object may go unnoticed until someone next asks for it. \emph{Role here:} the object store holds the data and is treated as untrusted.

\subsection{Cryptographic Hash Functions}
SHA-256, specified in FIPS 180-4~\cite{fips1804}, compresses an input of arbitrary length into a 256-bit value. The properties our argument needs are those formalized by Rogaway and Shrimpton~\cite{rogaway2004hash}: second-preimage resistance (given one input, finding another with the same output is infeasible) and collision resistance (finding any two inputs with the same output is infeasible). NIST rates the collision strength of SHA-256 at 128 bits~\cite{nist800107}. These guarantees are not universal across hash functions: a practical collision for SHA-1 was demonstrated in 2017~\cite{stevens2017sha1}, which is why the digest algorithm has to be chosen with care. \emph{Role here:} each object is reduced to a 32-byte digest, and the digest changes if any bit of the object changes. A set of $n$ digests, however, still has to be stored somewhere trustworthy; hashing by itself does not remove that need.

\subsection{Merkle Trees}
Merkle's hash tree~\cite{merkle1980protocols,merkle1987} places digests at the leaves of a binary tree and labels every internal node with the hash of its two children. Because the label at the top depends on every leaf and on their order, the root acts as a short \emph{commitment} to the whole sequence: altering, dropping or permuting a leaf alters the root. Building the tree costs exactly $n-1$ node hashes for $n$ leaves. The same structure underlies tamper-evident logs~\cite{crosby2009tamper}, the Certificate Transparency protocol~\cite{rfc6962,rfc9162} and file systems on untrusted servers~\cite{li2004sundr}. Certificate Transparency also introduced the practice, which we follow, of hashing leaves and internal nodes with different one-byte prefixes, so that an internal node can never be passed off as a leaf~\cite{rfc6962}. \emph{Role here:} the tree replaces $n$ trusted digests by one trusted 32-byte root.

\subsection{Merkle Proofs}
To show that a given leaf belongs to a tree with a known root, it suffices to supply the hashes of the siblings met on the way from the leaf to the root, often called the authentication or audit path. The verifier folds these siblings into the leaf, one level at a time, and checks whether the result equals the root. A balanced tree with $n$ leaves needs at most $\lceil \log_2 n \rceil$ siblings, so the proof is small regardless of how large the object is; Szydlo~\cite{szydlo2004traversal} studies how such paths can be produced efficiently for successive leaves. The path does not have to come from a trusted party: a wrong sibling simply leads to a wrong root. \emph{Role here:} one object can be checked without fetching the others, and the untrusted provider can serve the proofs.

\subsection{Blockchains and Smart Contracts}
A public blockchain is a ledger replicated across many nodes, extended only by appending blocks, and practically immutable unless an attacker controls a large share of the network's resources~\cite{nakamoto2008,zheng2017overview}. Ethereum~\cite{buterin2014,wood2014} lets programs, called smart contracts, keep state on this ledger. The state changes only through transactions that pass the contract's own checks and can be read by anyone. Execution is metered in \emph{gas}; writing a new 32-byte storage word is among the most expensive operations, so contracts are usually designed to keep large data off-chain and store only short commitments on-chain~\cite{zyskind2015decentralizing}. Contract code can also contain defects~\cite{atzei2017survey,luu2016smarter}, which is why our threat model treats the correctness of the contract as an explicit assumption. \emph{Role here:} the contract is a public, provider-independent place for the root. It never sees the objects and cannot itself notice tampering; detection happens only when an auditor compares freshly computed values with the stored root.

\section{Related Work}\label{sec:related}

We group prior work into four lines: proofs of storage, public and dynamic auditing, hash-tree structures for untrusted storage, and blockchain-assisted auditing. Table~\ref{tab:related} compares representative schemes along the dimensions that matter for our design.

\subsection{Provable Data Possession and Proofs of Retrievability}
Provable data possession (PDP), proposed by Ateniese et al.~\cite{ateniese2007pdp} and extended in a later journal version~\cite{ateniese2011remote}, lets a client check that a server still stores a file. The client attaches homomorphic tags to file blocks before upload; afterwards it challenges a random subset of blocks, and the server answers with one short aggregated value instead of the blocks themselves. A lighter variant relies only on symmetric primitives and allows a bounded number of updates~\cite{ateniese2008scalable}, and MR-PDP~\cite{curtmola2008mrpdp} extends the idea to several replicas of the same file. Proofs of retrievability (PoR) give a stronger guarantee: a server that passes the checks must hold enough data for the file to be recovered. Juels and Kaliski~\cite{juels2007por} obtain this with hidden sentinel blocks and error-correcting codes; Shacham and Waters~\cite{shacham2008compact} construct compact PoRs, one of which can be verified with public parameters; Dodis et al.~\cite{dodis2009hardness} analyze PoR through hardness amplification; and HAIL~\cite{bowers2009hail} spreads a file over several providers so that integrity and availability can be checked and restored together. All of these protocols operate on blocks inside a file and keep their verification material with the owner or in public keys rather than on a ledger. The idea we borrow from them is random sampling: challenging $c$ random units catches a corrupted fraction with a probability that can be computed in advance (Section~\ref{sec:sampling}).

\subsection{Public and Dynamic Auditing}
Public auditing delegates verification to any third party. Wang et al. set out the requirements for publicly auditable cloud storage~\cite{wang2010network} and designed auditing that hides the data from the auditor by masking the server's response~\cite{wang2010infocom,wang2013tc}. In a separate scheme~\cite{wang2011tpds}, they authenticate block tags with a Merkle hash tree, so that the auditor can check block positions and the owner can modify, insert or delete blocks; the owner signs each new root and the server keeps it. Erway et al.~\cite{erway2009dpdp} support updates with a rank-based authenticated skip list. Yang and Jia~\cite{yang2013dynamic} add dynamic updates and batch auditing over many owners and clouds, and Zhu et al.~\cite{zhu2012cooperative} combine responses from several providers in a multicloud setting. In this line of work, the auditor trusts the reference value because the owner has signed it.

\subsection{Hash Trees over Untrusted Storage}
Hash trees have long been used to protect data held by parties that are not trusted. SUNDR~\cite{li2004sundr} keeps a file system on an untrusted server while letting clients detect misbehavior, and Crosby and Wallach~\cite{crosby2009tamper} build tamper-evident logs from a history tree that yields compact membership and consistency proofs. Certificate Transparency~\cite{rfc6962,rfc9162} publishes certificates in Merkle-tree logs that anyone can audit. Our commitment follows the leaf and node encoding of these logs, but it is applied to the objects of a storage bucket and the root is anchored in a smart contract.

\subsection{Blockchain-Assisted Auditing}
Several schemes use a blockchain to remove or constrain the third-party auditor. Liu et al.~\cite{liu2017icws} let owners and consumers of IoT data verify integrity through smart contracts, and Yue et al.~\cite{yue2018icpads} combine Merkle trees with a blockchain for peer-to-peer cloud storage and study how many samples to check. Zhang et al.~\cite{zhang2019procrastinating} keep certificateless tags for the proof itself and use Ethereum to derive unpredictable challenges and to time-stamp each audit, so that a late auditor can be identified. Wang et al.~\cite{wang2020fairpay} replace the auditor with a contract that pays the provider only when its periodic possession proofs verify. Shu et al.~\cite{shu2022decentralized} decentralize auditing over a blockchain network and use a DAO-style design to limit dishonest auditors and miners. Lin et al.~\cite{lin2022consortium} record auditor activity on a consortium chain, where a contract flags abnormal behavior. These works study proof aggregation, incentives and auditor accountability; to our knowledge, they do not provide an S3-level, per-object classification of objects as intact, modified or missing together with a measured breakdown of where audit cost arises.

\subsection{Positioning}
Our design deliberately uses fewer primitives than the tag-based schemes: SHA-256 only, no keys for verification, and one 32-byte root per dataset on-chain. The price is that each audited object must be downloaded, so communication savings come from auditing a subset of objects rather than from aggregate proofs. We do not claim stronger security than the schemes above. Our contribution is an implementation and measurement study of the simpler design on an S3-compatible store, with per-object, three-way classification.

\begin{table}[t]
\centering
\caption{Comparison with Related Work}
\label{tab:related}
\setlength{\tabcolsep}{3pt}
\footnotesize
\begin{adjustbox}{max width=\columnwidth}
\begin{tabular}{@{}lllll@{}}
\toprule
Work & Integrity primitive & Merkle & Blockchain role & Public \\
\midrule
PDP~\cite{ateniese2007pdp} & homomorphic tags, sampling & no & none & yes \\
PoR~\cite{juels2007por} & sentinels + ECC & no & none & no \\
Compact PoR~\cite{shacham2008compact} & BLS / MAC authenticators & no & none & yes$^\dagger$ \\
DPDP~\cite{erway2009dpdp} & authenticated skip list & related & none & no \\
Wang et al.~\cite{wang2011tpds} & BLS tags + MHT & yes & none & yes \\
Liu et al.~\cite{liu2017icws} & data hashes, contracts & n/r & verification & yes \\
Yue et al.~\cite{yue2018icpads} & Merkle tree, sampling & yes & verification & yes \\
Zhang et al.~\cite{zhang2019procrastinating} & certificateless tags & no & challenges, log & yes \\
\textbf{This work} & SHA-256 Merkle tree & yes & root + audit log & yes \\
\bottomrule
\end{tabular}
\end{adjustbox}\\[2pt]
\raggedright\scriptsize $^\dagger$BLS variant. n/r: not reported in the sources we consulted.
\end{table}

\section{Problem Statement, Threat Model and Requirements}\label{sec:problem}

\subsection{Problem Statement}
A data owner stores a dataset $\mathcal{D} = \{(k_1, D_1), \ldots, (k_n, D_n)\}$ of objects with keys $k_i$ and contents $D_i$ in a bucket of an object store. At a later time, any party should be able to decide, for any object or for the whole dataset, whether the stored data still equals $\mathcal{D}$, and to identify which objects were modified or deleted, without trusting the provider, without the owner being online, and without downloading the dataset for every check.

\subsection{System Entities}
\begin{itemize}
\item \textbf{Data owner} uploads $\mathcal{D}$, computes the commitment and registers it on-chain. Trusted at commitment time.
\item \textbf{Cloud storage provider (CSP)} stores the objects and the off-chain tree metadata and runs the \emph{prover}, which answers proof requests. Untrusted for integrity.
\item \textbf{Public auditor} challenges objects, verifies proofs and may record the outcome on-chain. Needs no secret; anyone can act as auditor.
\item \textbf{Smart contract} on an Ethereum-compatible chain stores the trusted root, the object count and the audit log.
\end{itemize}

\subsection{Threat Model}
The CSP, or an attacker who controls it, may: (i) modify the content of one or more objects; (ii) delete objects; (iii) replace an object with another stored object's content or rename it; (iv) alter the off-chain tree metadata; (v) return forged sibling hashes, wrong keys or proofs for a different index; and (vi) answer audits inconsistently, for example serving correct data only to some requests. It may also try to change the integrity reference itself.

\textit{Assumptions.} SHA-256 is collision- and second-preimage-resistant. The chain provides integrity of committed state and the contract code is correct. The owner's signing key is not compromised, and the owner computes the commitment over the intended data. The auditor's own environment (client code, reading of chain state) is not compromised.

\textit{Out of scope.} The mechanism does not protect against: data that was already wrong when the root was computed; a compromised owner key or root-generation process, which can register a wrong root; denial of service and unavailability of the CSP or the chain; a compromised auditor client; loss of confidentiality (objects are not encrypted by this mechanism); and recovery of lost data, since auditing detects but does not repair.

\subsection{Requirements}
\begin{itemize}
\item[R1] \textit{Detection:} any modification or deletion of a committed object is detected when that object is audited.
\item[R2] \textit{Localization:} a full audit identifies exactly which objects failed and whether each was modified or missing.
\item[R3] \textit{Public verifiability:} verification uses only public information (the on-chain root and object count).
\item[R4] \textit{Tamper-resistant reference:} the CSP cannot change the trusted root.
\item[R5] \textit{Efficiency:} a single object can be audited without retrieving other objects, and on-chain state is independent of $n$.
\end{itemize}

\section{Proposed System}\label{sec:system}

\subsection{System Architecture}
\begin{figure}[t]
\centering
\begin{tikzpicture}[
  font=\scriptsize, >={Stealth[length=3.6pt,width=3pt]}, line width=0.55pt,
  act/.style={draw, rounded corners=1.5pt, align=center, minimum width=2.0cm, text width=1.88cm,
              minimum height=0.8cm, inner sep=1.5pt, fill=white},
  hdr/.style={font=\scriptsize\bfseries, align=center},
  lab/.style={font=\scriptsize, align=center, inner sep=1pt},
  com/.style={->, draw=black},
  aud/.style={->, draw=blue!70!black},
  opt/.style={->, draw=blue!70!black, dashed}]
\fill[gray!8, rounded corners=3pt] (-0.05,0.78) rectangle (8.55,-3.02);
\node[font=\scriptsize\itshape, anchor=north west, text=gray!55!black] at (-0.03,-2.72) {off-chain};
\fill[orange!10, rounded corners=3pt] (-0.05,-3.17) rectangle (8.55,-4.37);
\node[font=\scriptsize\itshape, anchor=north west, text=gray!55!black] at (-0.03,-3.19) {on-chain};
\node[hdr] at (1.05,0.48) {Data owner};
\node[hdr] at (4.25,0.48) {Provider (untrusted)};
\node[hdr] at (7.45,0.48) {Public auditor};
\node[act] (up)  at (1.05,-0.25) {Upload objects\\$D_1,\ldots,D_n$};
\node[act] (bld) at (1.05,-1.5)  {Hash $H_i$, leaf $\ell_i$;\\build tree, root $R$};
\node[act] (st)  at (4.25,-0.25) {Object store:\\objects $D_i$};
\node[act] (pr)  at (4.25,-1.5)  {Prover: tree\\metadata};
\node[act] (get) at (7.45,-0.25) {Download\\$D'_i=\mathrm{GET}(k_i)$};
\node[act] (ver) at (7.45,-1.5)  {Leaf $\ell'_i$; root $R'_i$\\from proof $\pi_i$};
\node[act] (dec) at (7.45,-2.6)  {Compare $R'_i$, $R^\ast$\\$\rightarrow$ status};
\node[act, text width=3.3cm, fill=orange!4] (sc) at (4.25,-3.77)
   {Smart contract: $R^\ast$, $n$,\\version, audit log};
\draw[com] (up) -- node[lab, above]{\textbf{C1}}node[lab, below]{PUT} (st);
\draw[com] (up) -- node[lab, right]{\textbf{C2}} (bld);
\draw[com] (bld) -- node[lab, above]{\textbf{C3}}node[lab, below]{tree} (pr);
\draw[com] (bld.south) |- node[lab, pos=0.32, right, align=left]{\textbf{C4}\\registerRoot} (sc.west);
\draw[aud] (st) -- node[lab, above]{\textbf{A3}}node[lab, below]{$D'_i$} (get);
\draw[aud, <->] (pr) -- node[lab, above]{\textbf{A2} $i$}node[lab, below]{$k_i,\pi_i$} (ver);
\draw[aud] (get) -- node[lab, right]{\textbf{A4}} (ver);
\draw[aud] (ver) -- node[lab, right]{\textbf{A5}} (dec);
\draw[aud] ([yshift=3pt]sc.east) -| node[lab, pos=0.72, left, align=right]{\textbf{A1}\\getRoot} ([xshift=-9pt]dec.south);
\draw[opt] ([xshift=9pt]dec.south) |- node[lab, pos=0.3, right, align=left]{\textbf{A6}\\record-\\Audit} ([yshift=-3pt]sc.east);
\end{tikzpicture}
\caption{Proposed system architecture. Black arrows (C1--C4) form the commitment phase: upload, hashing and tree construction, storage of the tree metadata with the provider, and registration of the root. Blue arrows (A1--A6) form the audit phase of Algorithm~\ref{alg:audit}: read the trusted root, obtain key and proof, download the object, recompute the root and classify the object; recording the outcome on-chain is optional (dashed). Only the root, the object count and audit records are stored on-chain.}
\label{fig:arch}
\end{figure}
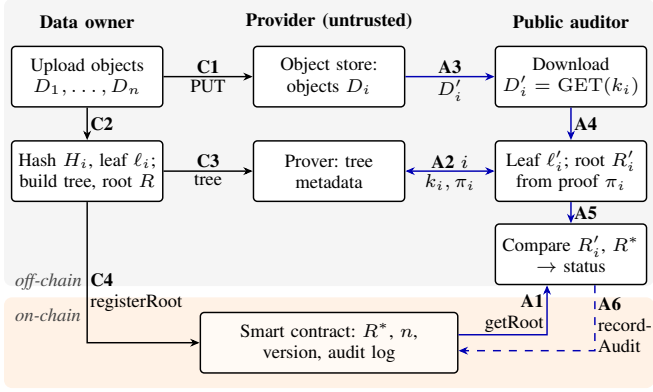

Fig.~\ref{fig:arch} shows the architecture and the two phases of the protocol. In the \emph{commitment phase}, the owner uploads the objects, hashes them, builds the Merkle tree and registers the root and the object count in the contract; the full tree (object keys and node hashes) is stored off-chain next to the data, where the prover can use it. In the \emph{audit phase}, the auditor reads the trusted root from the contract, chooses which objects to challenge, downloads each challenged object, obtains its sibling path from the prover, recomputes the root and compares. The division of state is strict:

\begin{itemize}
\item \textbf{Off-chain (object store):} object contents; object keys; the Merkle tree metadata ($n$ keys and $2n-1$ node hashes). None of it is trusted.
\item \textbf{On-chain (contract):} dataset identifier, Merkle root (32 bytes), object count $n$, registration time, version and owner address; one audit record per audit (root checked against, time, objects checked, objects failed, auditor). No object data.
\end{itemize}

\subsection{Data Upload and Hash Generation}
For each object the owner computes the content digest
\begin{equation}
H_i = \Hs(D_i), \qquad 1 \le i \le n, \label{eq:digest}
\end{equation}
and the leaf node
\begin{equation}
\ell_i = \Hs\big(\texttt{0x00} \concat \Hs(k_i) \concat H_i\big). \label{eq:leaf}
\end{equation}
Binding the key into the leaf means that an object moved to another key, or two objects swapped, no longer match their leaves; hashing the key keeps the leaf input at a fixed 65 bytes. Leaves are ordered lexicographically by key, which fixes each object's index $i$.

\subsection{Merkle Tree Construction}
\begin{figure}[t]
\centering
\begin{tikzpicture}[font=\scriptsize, line width=0.6pt, >={Stealth[length=3.6pt]},
  lf/.style={draw, rounded corners=1.5pt, minimum width=0.95cm, minimum height=0.48cm, inner sep=1pt, fill=white},
  inn/.style={draw, rounded corners=1.5pt, minimum width=1.25cm, minimum height=0.48cm, inner sep=1pt, fill=gray!8},
  pr/.style={draw, dashed, rounded corners=1.5pt, minimum width=0.95cm, minimum height=0.48cm, inner sep=1pt, fill=white},
  ob/.style={font=\scriptsize, align=center, inner sep=0.5pt},
  lev/.style={font=\scriptsize\itshape, anchor=east, text=gray!60!black}]
\foreach \i/\x in {1/0.75,2/2.05,3/3.35,4/4.65,5/6.85}
  \node[lf] (l\i) at (\x,0) {$\ell_{\i}$};
\foreach \i/\x in {1/0.75,2/2.05,3/3.35,4/4.65,5/6.85}
  { \node[ob] (o\i) at (\x,-0.78) {$(k_{\i},D_{\i})$}; \draw[->] (o\i) -- (l\i); }
\node[inn] (p12) at (1.4,1.05) {$P(\ell_1,\ell_2)$};
\node[inn] (p34) at (4.0,1.05) {$P(\ell_3,\ell_4)$};
\node[pr] (c5a) at (6.85,1.05) {$\ell_5$};
\node[inn] (p14) at (2.7,2.1) {$P(x_{12},x_{34})$};
\node[pr] (c5b) at (6.85,2.1) {$\ell_5$};
\node[inn, fill=blue!8, draw=blue!70!black, line width=0.8pt] (r) at (4.8,3.15) {$R = P(x_{1\text{--}4},\ell_5)$};
\draw (l1) -- (p12); \draw (l2) -- (p12); \draw (l3) -- (p34); \draw (l4) -- (p34);
\draw (p12) -- (p14); \draw (p34) -- (p14); \draw (p14) -- (r); \draw (c5b) -- (r);
\draw[->, dashed] (l5) -- node[ob, right]{promoted} (c5a);
\draw[->, dashed] (c5a) -- node[ob, right]{promoted} (c5b);
\node[lev] at (0.25,0) {$j{=}0$}; \node[lev] at (0.25,1.05) {$j{=}1$};
\node[lev] at (0.25,2.1) {$j{=}2$}; \node[lev] at (0.25,3.15) {$j{=}3$};
\node[ob, align=left, anchor=north west] at (-0.35,-1.12) {leaf: $\ell_i=\Hs(\texttt{0x00}\,\|\,\Hs(k_i)\,\|\,\Hs(D_i))$, Eq.~\eqref{eq:leaf}\\node: $P(x,y)=\Hs(\texttt{0x01}\,\|\,x\,\|\,y)$, Eq.~\eqref{eq:node}};
\end{tikzpicture}
\caption{Construction of the tree for an odd number of objects ($n=5$), following Eq.~\eqref{eq:level} and Algorithm~\ref{alg:build}. Shaded nodes are hashed with the \texttt{0x01} prefix; the unpaired leaf $\ell_5$ (dashed) is promoted unchanged at levels 0 and 1 instead of being paired with a copy of itself. The root needs $n-1=4$ node hashes, and the proof for $\ell_5$ contains a single sibling ($x_{1\text{--}4}$), illustrating that proofs can be shorter than $\lceil\log_2 n\rceil$.}
\label{fig:construct}
\end{figure}
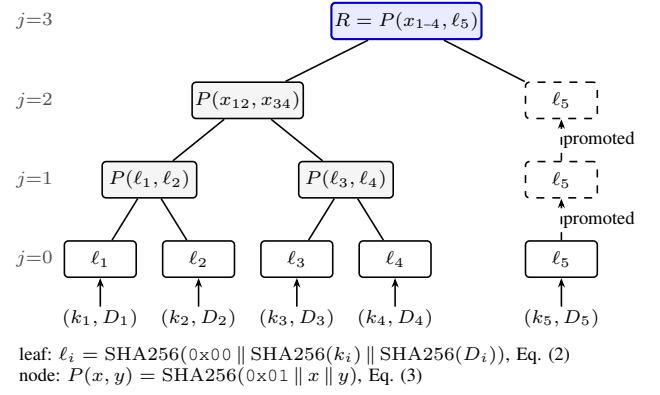

Internal nodes use a different prefix from leaves:
\begin{equation}
P(x_L, x_R) = \Hs\big(\texttt{0x01} \concat x_L \concat x_R\big). \label{eq:node}
\end{equation}
Let level $0$ be the leaves, $x^{(0)}_i = \ell_i$, and let $m_j$ be the number of nodes at level $j$. Level $j+1$ is
\begin{equation}
x^{(j+1)}_q =
\begin{cases}
P\big(x^{(j)}_{2q-1}, x^{(j)}_{2q}\big), & 2q \le m_j,\\
x^{(j)}_{2q-1}, & 2q-1 = m_j \text{ (unpaired)},
\end{cases}
\label{eq:level}
\end{equation}
for $1 \le q \le \lceil m_j/2 \rceil$. An unpaired last node is \emph{promoted} unchanged rather than hashed with a copy of itself, so no two different leaf sequences produce the same tree. Construction stops when one node remains:
\begin{equation}
R = \mathrm{MerkleRoot}(\ell_1, \ldots, \ell_n), \label{eq:root}
\end{equation}
after exactly $n-1$ evaluations of~\eqref{eq:node}. Algorithm~\ref{alg:build} gives the procedure, and Fig.~\ref{fig:construct} shows it for five objects, including the promotion of an unpaired node.

\begin{algorithm}[t]
\caption{Merkle Tree Construction}\label{alg:build}
\begin{algorithmic}[1]
\Require leaves $\ell_1,\ldots,\ell_n$ ordered by key
\Ensure levels $L_0,\ldots,L_d$; root $R$
\State $L_0 \gets [\ell_1,\ldots,\ell_n]$; $j \gets 0$
\While{$|L_j| > 1$}
  \State $L_{j+1} \gets [\,]$
  \For{$q \gets 1$ \textbf{to} $\lfloor |L_j|/2 \rfloor$}
    \State append $P(L_j[2q-1], L_j[2q])$ to $L_{j+1}$ \Comment{Eq.~\eqref{eq:node}}
  \EndFor
  \If{$|L_j|$ is odd} append $L_j[|L_j|]$ to $L_{j+1}$ \Comment{promote}
  \EndIf
  \State $j \gets j+1$
\EndWhile
\State \Return $L_0,\ldots,L_j$ and $R \gets L_j[1]$
\end{algorithmic}
\end{algorithm}

\subsection{Root Registration}
Algorithm~\ref{alg:commit} shows the commitment. The owner calls \texttt{registerRoot}$(\mathit{id}, R, n)$ with $R$ from Eq.~\eqref{eq:root}; the contract stores $R$, $n$, the block time, a version counter and the caller as owner, rejects calls from any other address for an existing $\mathit{id}$, and emits an event. A legitimate update of the dataset is committed by registering a new root, which increments the version. The tree metadata is written to the object store; it does not need protection because every value derived from it is checked against $R$.

\begin{algorithm}[t]
\caption{Commitment and Root Registration (owner)}\label{alg:commit}
\begin{algorithmic}[1]
\Require objects $(k_i, D_i)_{i=1}^{n}$, dataset id $\mathit{id}$
\For{$i \gets 1$ \textbf{to} $n$}
  \State PUT$(k_i, D_i)$; \ $H_i \gets \Hs(D_i)$; \ $\ell_i \gets$ Eq.~\eqref{eq:leaf}
\EndFor
\State $(L_0..L_d, R) \gets$ Algorithm~\ref{alg:build}$(\ell_1..\ell_n)$
\State store $(k_1..k_n, L_0..L_d)$ off-chain for the prover
\State send transaction \texttt{registerRoot}$(\mathit{id}, R, n)$; wait for receipt
\end{algorithmic}
\end{algorithm}

\subsection{Merkle Proof Generation and Verification}
For index $i$ the prover returns the key $k_i$ and the proof $\pi_i = (s_1,\ldots,s_m;\,b)$, where $s_j$ are the sibling hashes from the leaf upwards and bit $b_j$ records whether $s_j$ is a left sibling. Levels at which the path node was promoted contribute no sibling, so
\begin{equation}
m = |\pi_i| \le \lceil \log_2 n \rceil . \label{eq:prooflen}
\end{equation}
Verification (Algorithm~\ref{alg:verify}) folds the siblings into the leaf with~\eqref{eq:node} and returns the reconstructed root $R'_i$. It costs at most $\lceil \log_2 n\rceil$ hash evaluations, independent of the object size. The same function is implemented in Solidity as \texttt{verifyProof}, using the EVM's SHA-256 precompile.

\begin{algorithm}[t]
\caption{Merkle Proof Verification}\label{alg:verify}
\begin{algorithmic}[1]
\Require leaf $\ell$, proof $(s_1..s_m; b)$
\Ensure reconstructed root $R'$
\State $h \gets \ell$
\For{$j \gets 1$ \textbf{to} $m$}
  \If{$b_j = 1$} $h \gets P(s_j, h)$ \Else\ $h \gets P(h, s_j)$
  \EndIf
\EndFor
\State \Return $h$
\end{algorithmic}
\end{algorithm}

\begin{figure}[t]
\centering
\begin{tikzpicture}[font=\scriptsize, line width=0.6pt, >={Stealth[length=3.6pt]},
  hd/.style={draw, rounded corners=1.5pt, fill=gray!8, minimum width=1.75cm, minimum height=0.5cm, align=center, inner sep=1.5pt},
  msg/.style={->}, ret/.style={->, dashed},
  ml/.style={font=\scriptsize, fill=white, inner sep=0.8pt},
  act/.style={font=\scriptsize, align=left, anchor=west}]
\def\xo{0.85} \def\xs{3.2} \def\xc{5.55} \def\xa{7.75}
\node[hd] at (\xo,0) {Data owner};
\node[hd] at (\xs,0) {Provider\\(store + prover)};
\node[hd] at (\xc,0) {Smart\\contract};
\node[hd] at (\xa,0) {Public\\auditor};
\foreach \x in {\xo,\xs,\xc,\xa} \draw[gray!60, dashed] (\x,-0.35) -- (\x,-8.15);
\fill[gray!6] (0,-0.55) rectangle (8.6,-0.8);
\node[font=\scriptsize\bfseries] at (4.3,-0.675) {Commitment phase (Algorithm~\ref{alg:commit})};
\draw[msg] (\xo,-1.15) -- node[ml, above]{1: PUT$(k_i,D_i)$, $i=1..n$} (\xs,-1.15);
\draw[->] (\xo,-1.5) -- ++(0.35,0) |- (\xo,-1.95);
\node[act] at (\xo+0.4,-1.73) {2: $H_i,\ell_i$; tree, $R$};
\draw[msg] (\xo,-2.3) -- node[ml, above, pos=0.5]{3: tree metadata} (\xs,-2.3);
\draw[msg] (\xo,-2.75) -- node[ml, above, pos=0.62]{4: registerRoot$(id,R,n)$ [tx]} (\xc,-2.75);
\draw[ret] (\xc,-3.15) -- node[ml, above, pos=0.6]{5: receipt, RootRegistered} (\xo,-3.15);
\fill[blue!5] (0,-3.45) rectangle (8.6,-3.7);
\node[font=\scriptsize\bfseries] at (4.3,-3.575) {Audit phase (Algorithm~\ref{alg:audit})};
\draw[msg] (\xa,-4.05) -- node[ml, above]{6: getRoot$(id)$ [view]} (\xc,-4.05);
\draw[ret] (\xc,-4.45) -- node[ml, above]{7: $(R^\ast,n)$} (\xa,-4.45);
\draw[gray!70, rounded corners=2pt] (2.75,-4.75) rectangle (8.5,-7.15);
\node[font=\scriptsize, anchor=north west, fill=white, inner sep=1pt] at (2.8,-4.78) {\textbf{loop} for each $i\in C$};
\draw[msg] (\xa,-5.2) -- node[ml, above, pos=0.3]{8: challenge $i$} (\xs,-5.2);
\draw[ret] (\xs,-5.6) -- node[ml, above, pos=0.55]{9: $(k_i,\pi_i)$} (\xa,-5.6);
\draw[msg] (\xa,-6.0) -- node[ml, above, pos=0.55]{10: GET$(k_i)$} (\xs,-6.0);
\draw[ret] (\xs,-6.4) -- node[ml, above, pos=0.55]{11: $D'_i$ or \emph{no such key}} (\xa,-6.4);
\draw[->] (\xa,-6.6) -- ++(-0.3,0) |- (\xa,-6.95);
\node[act, anchor=east] at (\xa-0.35,-6.78) {12: $\ell'_i$, $R'_i$, status (Eq.~\eqref{eq:status})};
\draw[msg, dashed] (\xa,-7.55) -- node[ml, above]{13: recordAudit$(id,|C|,f)$ [tx, optional]} (\xc,-7.55);
\draw[ret] (\xc,-7.95) -- node[ml, above]{14: receipt} (\xa,-7.95);
\end{tikzpicture}
\caption{Message sequence of commitment registration and a subsequent audit. Messages 1--5 are sent once per dataset version; messages 6--14 form one audit over the challenge set $C$ (all $n$ indices, a random sample of $c$, or one index). Only messages 4 and 13 are blockchain transactions; message 6 is a read-only call. The provider supplies the proof in message 9, but the auditor downloads the object itself in messages 10--11.}
\label{fig:sequence}
\end{figure}
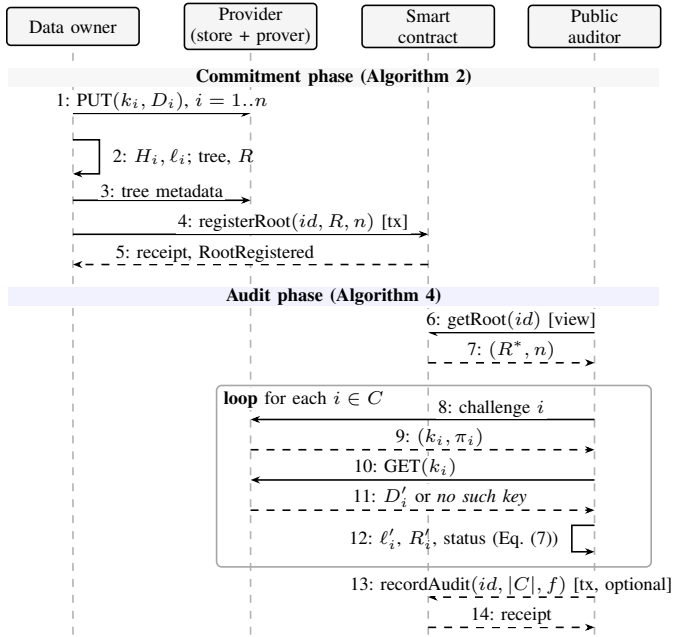

Fig.~\ref{fig:sequence} shows, as a message sequence, how this registration precedes every later audit and which steps involve the blockchain.

\subsection{Public Audit and Tampering/Deletion Detection}
The auditor reads $(R^\ast, n)$ from the contract and selects a challenge set $C \subseteq \{1..n\}$: all indices for a \emph{full audit}, $c$ indices drawn uniformly without replacement for a \emph{sampled audit}, or one index for a \emph{targeted audit}. For each $i \in C$ it obtains $(k_i, \pi_i)$ from the prover and downloads the object itself (Algorithm~\ref{alg:audit}). The status of object $i$ is
\begin{equation}
\mathrm{status}_i =
\begin{cases}
\textsc{missing}, & \text{GET}(k_i) \text{ returns no object},\\
\textsc{intact}, & R'_i = R^\ast,\\
\textsc{modified}, & R'_i \neq R^\ast,
\end{cases}
\label{eq:status}
\end{equation}
with $R'_i$ the root reconstructed from $\ell'_i = \Hs(\texttt{0x00}\concat\Hs(k_i)\concat\Hs(D'_i))$ and $\pi_i$, where $D'_i$ is the content currently stored. The audit passes iff every challenged object is intact.

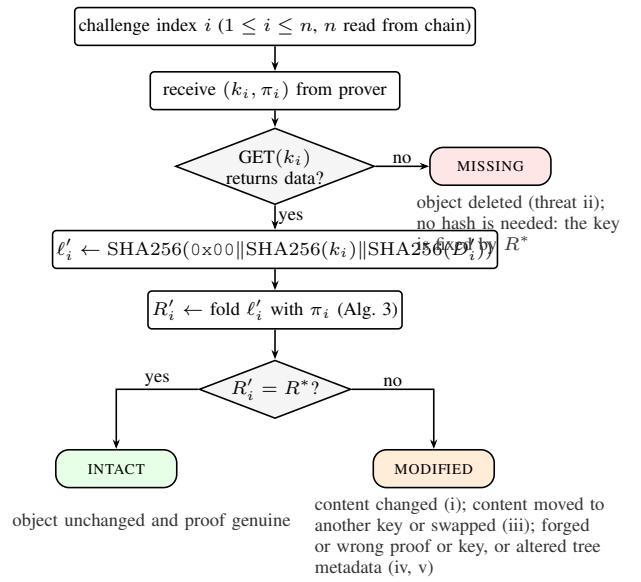
\begin{figure}[t]
\centering
\begin{tikzpicture}[font=\scriptsize, line width=0.6pt, >={Stealth[length=3.6pt]},
  st/.style={draw, rounded corners=1.5pt, align=center, minimum width=3.3cm, minimum height=0.5cm, inner sep=1.5pt, fill=white},
  dc/.style={draw, diamond, aspect=2.6, align=center, inner sep=0.5pt, fill=gray!8},
  oc/.style={draw, rounded corners=5pt, align=center, minimum width=1.6cm, minimum height=0.5cm, inner sep=1.5pt, font=\scriptsize\scshape},
  note/.style={font=\scriptsize, align=left, text=gray!40!black, anchor=west},
  lab/.style={font=\scriptsize, inner sep=1pt}]
\node[st] (s0) at (0,0) {challenge index $i$ ($1 \le i \le n$, $n$ read from chain)};
\node[st] (s1) at (0,-0.85) {receive $(k_i,\pi_i)$ from prover};
\node[dc] (d1) at (0,-1.85) {GET$(k_i)$\\returns data?};
\node[st] (s2) at (0,-2.95) {$\ell'_i \gets \Hs(\texttt{0x00}\|\Hs(k_i)\|\Hs(D'_i))$};
\node[st] (s3) at (0,-3.75) {$R'_i \gets$ fold $\ell'_i$ with $\pi_i$ (Alg.~\ref{alg:verify})};
\node[dc] (d2) at (0,-4.8) {$R'_i = R^\ast$?};
\node[oc, fill=green!10] (oi) at (-2.1,-5.85) {intact};
\node[oc, fill=orange!15] (om) at (2.1,-5.85) {modified};
\node[oc, fill=red!10] (ox) at (2.85,-1.85) {missing};
\draw[->] (s0) -- (s1); \draw[->] (s1) -- (d1);
\draw[->] (d1) -- node[lab, right]{yes} (s2);
\draw[->] (d1) -- node[lab, above]{no} (ox);
\draw[->] (s2) -- (s3); \draw[->] (s3) -- (d2);
\draw[->] (d2) -| node[lab, above, pos=0.25]{yes} (oi);
\draw[->] (d2) -| node[lab, above, pos=0.25]{no} (om);
\node[note, text width=2.7cm] at (1.75,-2.62) {object deleted (threat ii); no hash is needed: the key is fixed by $R^\ast$};
\node[note] at (-3.6,-6.55) {object unchanged and proof genuine};
\node[note, text width=4.1cm] at (0.4,-6.75) {content changed (i); content moved to another key or swapped (iii); forged or wrong proof or key, or altered tree metadata (iv, v)};
\end{tikzpicture}
\caption{Decision procedure that the auditor applies to each challenged object (Eq.~\eqref{eq:status}, Algorithm~\ref{alg:audit}), annotated with the threat-model actions of Section~\ref{sec:problem} that lead to each outcome. Deletion is recognized before any hashing; every other deviation, including misbehavior of the prover, ends in \textsc{modified}. The audit passes only if every challenged object is \textsc{intact}.}
\label{fig:decision}
\end{figure}

Downloading $D'_i$ rather than accepting a digest from the CSP is deliberate: a CSP that had lost or altered the object could otherwise keep returning the original digest. Deletion is detectable even though a deleted object cannot be hashed, because the key and index of every committed object are fixed by the tree and the count $n$ is on-chain: the auditor knows which object must exist, asks for it and receives none. If the prover instead omits or renames the entry, the returned key no longer reconstructs $R^\ast$ through~\eqref{eq:leaf}. An inconsistent prover is thus classified as \textsc{modified}, the conservative outcome. Fig.~\ref{fig:decision} summarizes this per-object decision and maps each attacker action of the threat model to the outcome it produces.

 The auditor may also publish the outcome with \texttt{recordAudit}$(\mathit{id}, |C|, \#\text{failed})$; the contract appends a time-stamped record that includes the root against which the audit was checked.

\begin{algorithm}[t]
\caption{Public Audit with Tampering/Deletion Detection}\label{alg:audit}
\begin{algorithmic}[1]
\Require dataset id $\mathit{id}$, challenge set $C$
\Ensure status of every $i \in C$; PASS/FAIL
\State $(R^\ast, n) \gets$ \texttt{getRoot}$(\mathit{id})$ \Comment{on-chain read}
\For{\textbf{each} $i \in C$}
  \State $(k_i, \pi_i) \gets$ prover.challenge$(i)$
  \State $D'_i \gets$ GET$(k_i)$
  \If{$D'_i = \bot$} $\mathrm{status}_i \gets \textsc{missing}$; \textbf{continue}
  \EndIf
  \State $\ell'_i \gets \Hs(\texttt{0x00}\concat\Hs(k_i)\concat\Hs(D'_i))$
  \State $R'_i \gets$ Algorithm~\ref{alg:verify}$(\ell'_i, \pi_i)$
  \State $\mathrm{status}_i \gets \textsc{intact}$ \textbf{if} $R'_i = R^\ast$ \textbf{else} \textsc{modified}
\EndFor
\State \textbf{optional:} \texttt{recordAudit}$(\mathit{id}, |C|, |\{i: \mathrm{status}_i \ne \textsc{intact}\}|)$
\State \Return PASS iff all statuses are \textsc{intact}
\end{algorithmic}
\end{algorithm}

\subsection{Sampled Audits}\label{sec:sampling}
A full audit must download every object. A sampled audit of $c$ distinct objects, drawn uniformly from $n$ of which $t$ are modified or missing, detects at least one violation with probability
\begin{equation}
P_{\mathrm{det}}(n,t,c) = 1 - \binom{n-t}{c}\Big/\binom{n}{c}, \label{eq:pdet}
\end{equation}
the hypergeometric argument used in PDP~\cite{ateniese2007pdp,ateniese2011remote}. For a fixed fraction $t/n$, $P_{\mathrm{det}}$ depends mainly on $c$ and hardly on $n$, so the cost of reaching a target detection probability is roughly constant as the dataset grows.

\subsection{Communication Model}
A full-data audit transfers $B_{\mathrm{full}} = \sum_{i=1}^{n} |D_i|$ bytes and requires the auditor to keep $32n$ bytes of reference digests. A Merkle audit with challenge set $C$ transfers
\begin{equation}
B_{\mathrm{Merkle}}(C) = \sum_{i \in C} \big(|D_i| + 32\,|\pi_i| + 32 + |k_i|\big), \label{eq:bytes}
\end{equation}
where the 32-byte term is the direction bitmap as passed to the contract, and the auditor keeps no per-object state. For a full audit, $B_{\mathrm{Merkle}}$ slightly \emph{exceeds} $B_{\mathrm{full}}$; the saving comes only from auditing $|C| < n$ objects.

\section{Implementation}\label{sec:impl}

The prototype is written in Python and Solidity; Table~\ref{tab:components} lists its components. In total it comprises about 1{,}030 non-blank lines, including the experiment harness and plotting code.

\begin{table}[t]
\centering
\caption{Implementation Components (Non-Blank Lines)}
\label{tab:components}
\footnotesize
\setlength{\tabcolsep}{4pt}
\begin{adjustbox}{max width=\columnwidth}
\begin{tabular}{@{}llr@{}}
\toprule
Component & Responsibility & Lines \\
\midrule
\texttt{merkle.py} & digests, leaves, Alg.~\ref{alg:build}, proofs, Alg.~\ref{alg:verify} & 69 \\
\texttt{audit.py} & owner commit, prover, Alg.~\ref{alg:audit}, baseline & 136 \\
\texttt{storage.py} & S3 client wrapper (PUT/GET/DELETE/LIST) & 34 \\
\texttt{chain.py} & Web3.py deploy, register, read, record, verify & 66 \\
\texttt{MerkleAuditRegistry.sol} & registry contract + evaluation baseline & 87 \\
\texttt{compile.js} & solc compilation to ABI/bytecode & 13 \\
\texttt{experiments.py} & experiment harness, raw JSON records & 411 \\
\texttt{plots.py} & figures from raw records & 172 \\
\texttt{tests/} & unit and on-chain cross-check tests & 46 \\
\bottomrule
\end{tabular}
\end{adjustbox}
\end{table}

\textit{Merkle library.} Digests use Python's \texttt{hashlib} SHA-256. The tree is kept as a list of levels, so proof generation is a walk of $\lceil\log_2 n\rceil$ list lookups. Proofs carry the sibling list and an integer bitmap of directions, the same encoding the contract accepts.

\textit{Storage and prover.} Objects are stored in one bucket per dataset under flat keys (\texttt{obj-0000000.bin}, \ldots) and accessed with the MinIO Python SDK~\cite{minio}. The prover is a class that holds the commitment's levels and keys and answers \texttt{challenge}$(i)$ with $(k_i, \pi_i)$; in the experiments it runs in the auditor's process, so its time is measured separately and transport of the proof is accounted for by its byte size, Eq.~\eqref{eq:bytes}.

\textit{Smart contract.} \texttt{MerkleAuditRegistry} exposes \texttt{registerRoot}, \texttt{getRoot}, \texttt{recordAudit}, \texttt{getAudit}, \texttt{auditCount} and the view function \texttt{verifyProof}. A dataset occupies three storage slots (root; count, time and version; owner) and an audit record two. A second contract, \texttt{PerObjectHashRegistry}, which stores every object digest on-chain, exists only as an evaluation baseline. Both are compiled with solc~0.8.24 (optimizer, 200 runs; 2{,}656 bytes of runtime code for the registry) ~\cite{soliditydocs} and deployed on a local Hardhat network~\cite{hardhat}; \texttt{chain.py} calls them through Web3.py~\cite{web3py}.

\textit{Measurement.} Each audit records separately the time to read the root from the chain, to retrieve objects, to compute digests and leaves, to generate proofs and to verify them, using \texttt{time.perf\_counter}, as well as the bytes of object content and of proof metadata. Python's garbage collector is run before each repetition, and peak memory of commitment construction is measured with \texttt{tracemalloc} in a separate, untimed pass. For transactions we record gas used from the receipt and wall-clock time from submission to receipt; for view functions we record \texttt{eth\_estimateGas}, which includes the 21{,}000 intrinsic gas of a transaction, although a view call made off-chain costs nothing.

\textit{Validation.} Six automated tests check that every proof verifies for all indices of trees with 1--69, 127--129 and 1{,}000 leaves, that proof length never exceeds $\lceil\log_2 n\rceil$, that a modified leaf, a swapped key or a forged sibling fails, that the promotion rule of Eq.~\eqref{eq:level} holds, and that the Solidity \texttt{verifyProof} accepts every proof the Python verifier accepts (trees of 1, 2, 3, 7 and 33 leaves) and rejects a forged leaf.

\textit{A measurement artifact we removed.} In initial runs every GET took about 44~ms regardless of object size, while a HEAD request took 1~ms. The cause was Nagle's algorithm on the server socket interacting with delayed acknowledgements on the client. We patched the S3 server to set \texttt{TCP\_NODELAY}, after which a GET of 100~B to 200~KiB took 1.1--3.5~ms. All reported results use the patched server.

\section{Experimental Methodology}\label{sec:method}

\subsection{Experimental Environment}
All components run on one virtual machine (Table~\ref{tab:env}) and communicate over the loopback interface. The official MinIO server binary could not be obtained in our build environment, so the object store is s3s-fs~0.17.0~\cite{s3s}, an open-source S3-compatible server written in Rust that stores each object as a file; the client is the MinIO Python SDK, so the client side is the same as against MinIO. The blockchain is a Hardhat~\cite{hardhat} network with automatic mining, i.e. every transaction is mined immediately in its own block.

\begin{table}[t]
\centering
\caption{Experimental Environment}
\label{tab:env}
\footnotesize
\setlength{\tabcolsep}{3pt}
\begin{adjustbox}{max width=\columnwidth}
\begin{tabular}{@{}ll@{}}
\toprule
Component & Technology / configuration \\
\midrule
Hardware & Intel Xeon @ 2.10 GHz, 2 vCPU, 7.8 GiB RAM \\
Operating system & Ubuntu 24.04.5 LTS, Linux 6.18, ext4 \\
Programming language & Python 3.13.16 \\
Hash function & SHA-256 (Python \texttt{hashlib}) \\
Merkle tree & own implementation, Section~\ref{sec:system} \\
Object storage & s3s-fs 0.17.0 (S3 API, file backend), loopback \\
Storage client & MinIO Python SDK 7.2.20 \\
Blockchain & Hardhat 2.22.17 network, chain id 31337, \\
 & Cancun rules, automine, Node.js 22.22.0 \\
Smart contract language & Solidity 0.8.24, optimizer 200 runs \\
Blockchain interface & Web3.py 8.0.0 \\
Logging & per-run JSON records \\
\bottomrule
\end{tabular}
\end{adjustbox}
\end{table}

\subsection{Dataset Configuration}
Datasets are synthetic and reproducible: object $i$ of the dataset with $n$ objects has a size drawn log-uniformly between 1~KiB and 256~KiB and pseudo-random content, both from a seeded generator, so that a dataset can be regenerated byte for byte. Object contents are incompressible and their sizes span typical small documents and images. Table~\ref{tab:datasets} lists the four datasets. The \emph{Large} dataset (1{,}000 objects) is the subject of experiments E1--E4.

\begin{table}[t]
\centering
\caption{Dataset Configuration}
\label{tab:datasets}
\footnotesize
\setlength{\tabcolsep}{3.5pt}
\begin{adjustbox}{max width=\columnwidth}
\begin{tabular}{@{}lrrrrr@{}}
\toprule
Dataset & Objects & Total size & Mean object & Depth & Tree metadata \\
\midrule
Small & 10 & 0.43 MiB & 43.9 KiB & 4 & 822 B \\
Medium & 100 & 4.00 MiB & 41.0 KiB & 7 & 7.8 KiB \\
Large & 1,000 & 48.68 MiB & 49.9 KiB & 10 & 77.2 KiB \\
Very large & 10,000 & 441.66 MiB & 45.2 KiB & 14 & 771.6 KiB \\
\bottomrule
\end{tabular}
\end{adjustbox}\\[2pt]
\raggedright\scriptsize Depth $=\lceil\log_2 n\rceil$. Tree metadata: object keys plus all $2n-1$ node hashes.
\end{table}

In addition, a component benchmark builds trees from $2^4$ to $2^{20}$ (1{,}048{,}576) random in-memory leaves to isolate the cost of the Merkle operations from storage access.

\subsection{Metrics}
\begin{itemize}
\item \textit{Hash time:} SHA-256 of object contents plus leaf computation, Eqs.~\eqref{eq:digest}--\eqref{eq:leaf}.
\item \textit{Construction time:} Algorithm~\ref{alg:build} over existing leaves.
\item \textit{Proof generation / verification time:} per proof, and summed over an audit.
\item \textit{Chain read time:} \texttt{getRoot} call through JSON-RPC.
\item \textit{Retrieval time:} GET of the challenged objects.
\item \textit{Audit time:} wall-clock time of Algorithm~\ref{alg:audit}, including the chain read.
\item \textit{Detection outcome:} per-object status compared with the ground truth of the injected fault; detection rate = correctly classified faulty objects / faulty objects; false positives = intact objects classified otherwise.
\item \textit{Bytes transferred:} object content plus proof metadata, Eq.~\eqref{eq:bytes}.
\item \textit{Memory:} peak traced allocation during commitment construction.
\item \textit{Gas and transaction latency:} from receipts; storage slots from the contract layout.
\end{itemize}

\subsection{Experimental Procedure}
Each experiment starts from a trusted state: the dataset is uploaded, the owner computes the commitment from its local copy (Algorithm~\ref{alg:commit}) and the root is registered on a freshly deployed contract. Faults are injected by writing to the object store directly, as a misbehaving provider would: \emph{bit flip} (one random bit of the content inverted), \emph{append} (one byte added) and \emph{replace} (content replaced by random bytes of equal length) for modification, and DELETE for deletion. The modified or deleted objects are chosen uniformly at random with a fixed seed. After each repetition the affected objects are restored from the owner's copy, so that every repetition starts from the trusted state. Every configuration is repeated five times (three times for the 10{,}000-object scalability runs); we report the mean and the sample standard deviation over repetitions. The full-data baseline downloads every object and compares its SHA-256 digest with a digest list held by the auditor. All raw records, the code that produced them, and the scripts that generate every figure and table are part of the project artifact.

\section{Experimental Evaluation}\label{sec:results}

\subsection{E1: Normal Data Integrity Audit (RQ1)}
E1 establishes the trusted state of the Large dataset and audits it unchanged with a full audit ($C = \{1..n\}$) and, for comparison, with a whole-root recomputation from all current objects. Table~\ref{tab:e1} gives the measurements.

\begin{table}[t]
\centering
\caption{E1: Normal Integrity Audit (Large Dataset, $n = 1{,}000$)}
\label{tab:e1}
\footnotesize
\setlength{\tabcolsep}{4pt}
\begin{adjustbox}{max width=\columnwidth}
\begin{tabular}{@{}lr@{}}
\toprule
Quantity & Value \\
\midrule
\multicolumn{2}{@{}l}{\textit{Commitment (owner, one run at upload)}} \\
Objects / total size & 1,000 / 48.68 MiB \\
SHA-256 digests + leaves & 53.1 ms \\
Tree construction (2,001 nodes, depth 10) & 0.58 ms \\
Trusted root (prefix) & \texttt{107037f0dda3709a\ldots} \\
\texttt{registerRoot} & 91,863 gas, 11.1 ms \\
\midrule
\multicolumn{2}{@{}l}{\textit{Full audit (mean $\pm$ s.d., 5 runs)}} \\
Chain read (\texttt{getRoot}) & 10.4 $\pm$ 2.9 ms \\
Object retrieval (1,000 GETs) & 1,312.5 $\pm$ 109.8 ms \\
SHA-256 digests + leaves & 49.8 $\pm$ 1.8 ms \\
Proof generation (1,000 proofs) & 7.0 $\pm$ 0.4 ms \\
Proof verification (1,000 proofs) & 10.2 $\pm$ 0.5 ms \\
Total audit time & 1,393.8 $\pm$ 112.7 ms \\
Whole-root recomputation & 1,317.6 $\pm$ 75.3 ms \\
Result & 1,000/1,000 \textsc{intact}, PASS in 5/5 \\
\texttt{recordAudit} & 118,104 gas \\
\bottomrule
\end{tabular}
\end{adjustbox}
\end{table}

All 1{,}000 objects were classified as intact and the recomputed root equalled the registered root in every run, so the mechanism produces no false alarms on unmodified data. Retrieval accounts for 94\% of the audit time; hashing 48.68~MiB takes 3.6\%, and generating and verifying 1{,}000 proofs together 1.2\% (7.0~$\mu$s and 10.2~$\mu$s per proof). The cost of a full audit is therefore the cost of reading the data, not of the Merkle operations.

\finding{1}{On unmodified data every object passes and the audit time is dominated (94\%) by object retrieval; Merkle proof generation and verification take about 17~$\mu$s per object.}

\subsection{E2: Single-File Tampering (RQ1)}
\begin{figure}[t]
\centering
\begin{tikzpicture}[font=\scriptsize, line width=0.6pt, >={Stealth[length=3.6pt]},
  n/.style={draw, rounded corners=2pt, minimum width=0.78cm, minimum height=0.46cm, inner sep=1pt, fill=white},
  bad/.style={n, fill=orange!25, draw=orange!75!black, line width=1.1pt},
  sib/.style={n, fill=blue!10, draw=blue!70!black, line width=1.1pt, dashed}]
\foreach \i/\s/\t in {1/n/{\ell_1},2/n/{\ell_2},3/sib/{\ell_3},4/bad/{\ell'_4},5/n/{\ell_5},6/n/{\ell_6},7/n/{\ell_7},8/n/{\ell_8}}
  \node[\s] (l\i) at ({(\i-1)*0.9},0) {$\t$};
\node[sib] (a1) at (0.45,1.0) {$P_{12}$};
\node[bad] (a2) at (2.25,1.0) {$P'_{34}$};
\node[n]   (a3) at (4.05,1.0) {$P_{56}$};
\node[n]   (a4) at (5.85,1.0) {$P_{78}$};
\node[bad] (b1) at (1.35,2.0) {$P'_{1\text{--}4}$};
\node[sib] (b2) at (4.95,2.0) {$P_{5\text{--}8}$};
\node[bad] (r)  at (3.15,3.0) {$R'$};
\foreach \c/\p in {l1/a1,l2/a1,l3/a2,l4/a2,l5/a3,l6/a3,l7/a4,l8/a4,a1/b1,a2/b1,a3/b2,a4/b2,b1/r,b2/r}
  \draw (\c) -- (\p);
\node[draw, rounded corners=1pt, fill=orange!8, align=center] (f) at (2.7,-1.0) {$D'_4$ (modified object)};
\draw[->] (f) -- node[right, align=left]{SHA-256,\\Eq.~\eqref{eq:leaf}} (l4);
\node[draw, rounded corners=1pt, align=center, fill=white] (cmp) at (5.75,3.0) {$R' \neq R^\ast$ (on-chain)\\$\Rightarrow$ \textsc{modified}};
\draw[->] (r) -- (cmp);
\node[blue!60!black] at ($(l3)+(0,-0.42)$) {$s_1$};
\node[blue!60!black, anchor=east] at ($(a1.west)+(-0.04,0)$) {$s_2$};
\node[blue!60!black, anchor=west] at ($(b2.east)+(0.04,0)$) {$s_3$};
\node[anchor=west, align=left] at (4.3,-0.95) {\tikz\node[bad,minimum width=0.3cm,minimum height=0.2cm]{}; recomputed by the auditor\\[1pt]\tikz\node[sib,minimum width=0.3cm,minimum height=0.2cm]{}; sibling $s_j$ supplied in $\pi_4$};
\end{tikzpicture}
\caption{Effect of modifying object 4 in an 8-object tree (illustration, not measured data). The auditor hashes the downloaded content $D'_4$ into a new leaf $\ell'_4$ and recomputes every ancestor on its path (orange, solid border) using the three siblings $s_1$--$s_3$ supplied unchanged in the proof $\pi_4$ (blue, dashed border). The reconstructed root $R'$ differs from the registered root $R^\ast$, so object 4 is classified as \textsc{modified}.}
\label{fig:path}
\end{figure}
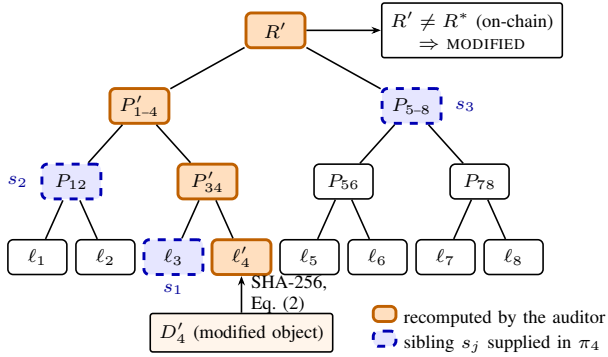

After establishing the trusted state, E2 modifies exactly one randomly chosen object, audits it with a targeted single-object audit, then runs a full audit and a whole-root recomputation, and restores the object. This is done five times for each of the three modification types. Table~\ref{tab:e2} reports the outcome and the cost of detection; Fig.~\ref{fig:path} illustrates the mechanism.

\begin{table}[t]
\centering
\caption{E2: Single-File Tampering ($n = 1{,}000$, 5 runs per type)}
\label{tab:e2}
\footnotesize
\setlength{\tabcolsep}{3pt}
\begin{adjustbox}{max width=\columnwidth}
\begin{tabular}{@{}lrrr@{}}
\toprule
 & Bit flip & Append 1 B & Replace \\
\midrule
Targeted audit: detected & 5/5 & 5/5 & 5/5 \\
Full audit: flagged objects & exactly 1 & exactly 1 & exactly 1 \\
Whole-root check fails & 5/5 & 5/5 & 5/5 \\
\midrule
Detection time (ms) & 9.77 $\pm$ 1.61 & 11.48 $\pm$ 2.16 & 11.37 $\pm$ 1.98 \\
\ \ chain read (ms) & 8.22 $\pm$ 0.98 & 9.99 $\pm$ 1.97 & 9.87 $\pm$ 1.98 \\
\ \ retrieval (ms) & 1.47 $\pm$ 0.79 & 1.43 $\pm$ 0.29 & 1.44 $\pm$ 0.27 \\
\ \ hashing ($\mu$s) & 28 $\pm$ 18 & 20 $\pm$ 9 & 23 $\pm$ 8 \\
\ \ verification ($\mu$s) & 28 $\pm$ 28 & 13 $\pm$ 2 & 13 $\pm$ 1 \\
Full audit time (s) & 1.53 $\pm$ 0.09 & 1.46 $\pm$ 0.06 & 1.44 $\pm$ 0.07 \\
\bottomrule
\end{tabular}
\end{adjustbox}
\end{table}

In all 15 trials the targeted audit returned \textsc{modified} for the tampered object, the full audit flagged that object and no other (999 intact), and the whole-root recomputation failed. For example, in the first bit-flip trial object \texttt{obj-0000121.bin} (4{,}339~B) had digest \texttt{966638ee\ldots} before and \texttt{351a68de\ldots} after the flip; folding the new leaf with its ten siblings gave \texttt{1dee077d\ldots}, which differs from the registered \texttt{107037f0\ldots}, and the whole-root recomputation produced the same \texttt{1dee077d\ldots}. The mechanism is that of Fig.~\ref{fig:path}: the changed digest changes the leaf~\eqref{eq:leaf}, and because each node on the path is a hash of its child, every ancestor up to the root changes, while the siblings supplied by the prover stay valid. The size of the change is irrelevant: a single bit and a full replacement are detected alike, as expected from a hash function. Detection takes about 10~ms, of which 8--10~ms is the JSON-RPC read of the root from the local node; retrieval of the object takes about 1.4~ms and hashing plus verification less than 0.06~ms.

\finding{2}{A single modified object, from a one-bit flip to full replacement, was detected in 15/15 trials and localized without false positives. A targeted audit takes about 10~ms, dominated by the on-chain root read.}

\subsection{E3: Multiple-File Tampering (RQ2)}
\begin{figure}[t]
\centering
\includegraphics[width=\columnwidth]{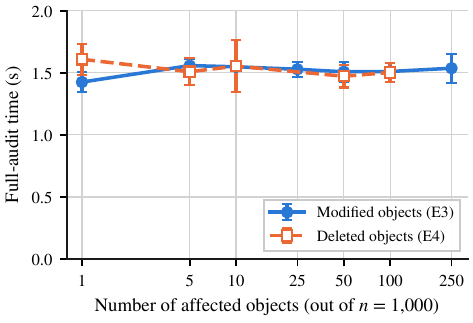}
\caption{Full audit time against the number of modified (E3) or deleted (E4) objects in the Large dataset (mean $\pm$ s.d., 5 runs).}
\label{fig:affected}
\end{figure}

E3 flips one bit in $t \in \{1, 5, 10, 25, 50, 100, 250\}$ randomly chosen objects of the Large dataset (0.1\%--25\%) and runs a full audit; five repetitions per level with different objects. Table~\ref{tab:e3} and Fig.~\ref{fig:affected} show the results.

\begin{table}[t]
\centering
\caption{E3: Multiple-File Tampering ($n = 1{,}000$, 5 runs per level)}
\label{tab:e3}
\footnotesize
\setlength{\tabcolsep}{3.5pt}
\begin{adjustbox}{max width=\columnwidth}
\begin{tabular}{@{}rrrrrr@{}}
\toprule
$t$ & Detected & False pos. & Detection rate & Verification (ms) & Audit time (s) \\
\midrule
1 & 5/5 & 0 & 100\% & 10.6 & 1.43 $\pm$ 0.08 \\
5 & 25/25 & 0 & 100\% & 11.5 & 1.56 $\pm$ 0.05 \\
10 & 50/50 & 0 & 100\% & 11.2 & 1.55 $\pm$ 0.02 \\
25 & 125/125 & 0 & 100\% & 10.6 & 1.53 $\pm$ 0.06 \\
50 & 250/250 & 0 & 100\% & 10.8 & 1.51 $\pm$ 0.08 \\
100 & 500/500 & 0 & 100\% & 11.4 & 1.51 $\pm$ 0.04 \\
250 & 1,250/1,250 & 0 & 100\% & 11.2 & 1.54 $\pm$ 0.12 \\
\bottomrule
\end{tabular}
\end{adjustbox}\\[2pt]
\raggedright\scriptsize Detected and false positives summed over 5 runs; verification = all 1,000 proofs of one audit (mean).
\end{table}

All 2{,}205 modified objects were classified as modified and none of the 32{,}795 untouched objects was flagged, a measured detection rate of 100\% with no false positives. Audit time does not depend on $t$ (1.43--1.56~s, within run-to-run variation), because a full audit performs the same retrieval, hashing and verification for every object whatever its status. Each object is checked against the root through its own path, so violations do not mask each other.

\begin{figure}[t]
\centering
\includegraphics[width=\columnwidth]{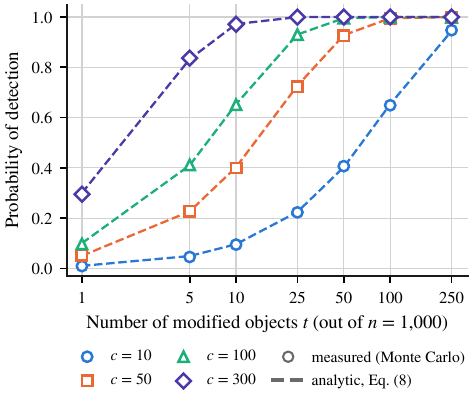}
\caption{Detection probability of a sampled audit of $c$ objects against the number of modified objects $t$ ($n = 1{,}000$). Markers: frequency over 5 tampered states $\times$ 2{,}000 random samples; dashed lines: Eq.~\eqref{eq:pdet}.}
\label{fig:sampling}
\end{figure}

\textit{Sampled audits.} Fig.~\ref{fig:sampling} shows the probability that a sampled audit of $c$ objects detects at least one of the $t$ modified objects. For every tampered state we used the per-object outcomes of the full audit and drew 2{,}000 random challenge sets of each size; the measured frequencies agree with Eq.~\eqref{eq:pdet} to within 0.005 in every cell (e.g., $t = 10$, $c = 100$: 0.651 measured, 0.653 analytic). In addition, real sampled audits were run once per repetition; they took 21.5--26.4~ms for $c = 10$, 74.8--90.3~ms for $c = 50$, 146.6--164.2~ms for $c = 100$ and 429.6--479.7~ms for $c = 300$. A sample of 300 objects (30\% of the work of a full audit) detected 1\% corruption ($t = 10$) with probability 0.97, whereas a single modified object is caught only with probability $c/n$.

\finding{3}{A full audit identified all 2{,}205 modified objects with no false positives at a cost independent of the number of violations. A sampled audit trades certainty for cost exactly as Eq.~\eqref{eq:pdet} predicts.}

\subsection{E4: File Deletion (RQ2)}
E4 deletes $d \in \{1, 5, 10, 50, 100\}$ random objects of the Large dataset and runs a targeted audit of one deleted object and a full audit (Table~\ref{tab:e4}). A mixed scenario then modifies 20 and deletes 20 objects at once to test the three-way classification.

\begin{table}[t]
\centering
\caption{E4: File Deletion ($n = 1{,}000$, 5 runs per level)}
\label{tab:e4}
\footnotesize
\setlength{\tabcolsep}{3pt}
\begin{adjustbox}{max width=\columnwidth}
\begin{tabular}{@{}rrrrr@{}}
\toprule
$d$ & \textsc{missing} found & Misclassified & Detection time (ms) & Full audit (s) \\
\midrule
1 & 5/5 & 0 & 10.43 $\pm$ 1.32 & 1.61 $\pm$ 0.13 \\
5 & 25/25 & 0 & 10.35 $\pm$ 0.92 & 1.51 $\pm$ 0.11 \\
10 & 50/50 & 0 & 10.76 $\pm$ 2.05 & 1.55 $\pm$ 0.21 \\
50 & 250/250 & 0 & 10.20 $\pm$ 1.94 & 1.47 $\pm$ 0.09 \\
100 & 500/500 & 0 & 10.89 $\pm$ 2.80 & 1.50 $\pm$ 0.08 \\
\midrule
\multicolumn{5}{@{}l}{Mixed, 20 modified + 20 deleted, 5 runs (classified as):} \\
\multicolumn{5}{@{}l}{\quad intact$\to$\textsc{intact} 4,800; modified$\to$\textsc{modified} 100; deleted$\to$\textsc{missing} 100} \\
\bottomrule
\end{tabular}
\end{adjustbox}
\end{table}

All 830 deleted objects were reported as \textsc{missing}, none as \textsc{modified}, and no present object as missing; the whole-root recomputation failed in every run. In the mixed scenario the classification matched the ground truth for all 5{,}000 object audits, so the audit distinguishes the three states. Deletion is detected without hashing the deleted object: the committed key list and the on-chain count $n$ say which objects must exist, and the auditor asks for each of them. A provider that removed the entry from the tree metadata would have to present a different key or proof for that index, which fails Eq.~\eqref{eq:status} as \textsc{modified}. Merely listing the bucket would not suffice, because the list shows what exists, not what should. Detection time (about 10~ms) and full-audit time match E2 and E3; the transfer saved by not downloading deleted objects is within run-to-run variation.

\finding{4}{All 830 deletions were detected and reported as \textsc{missing}, separately from modifications; deletion is detectable because the expected key set is fixed by the commitment, not by the current bucket listing.}

\subsection{E5: Scalability (RQ3)}
\begin{figure}[t]
\centering
\includegraphics[width=\columnwidth]{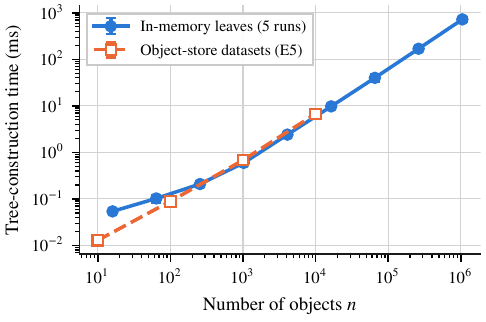}
\caption{Merkle tree construction time against the number of objects (log--log; mean $\pm$ s.d.): object-store datasets (E5) and in-memory component benchmark.}
\label{fig:build}
\end{figure}

\begin{figure}[t]
\centering
\includegraphics[width=\columnwidth]{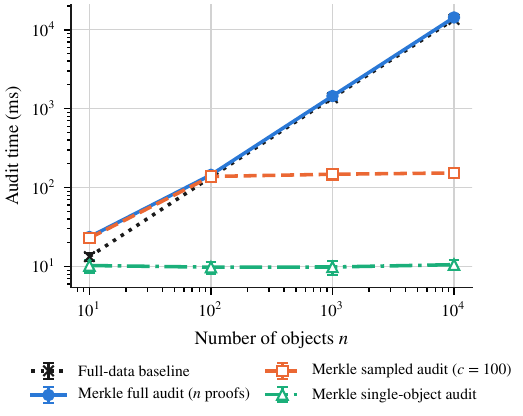}
\caption{Audit time against the number of objects (log--log; mean $\pm$ s.d.). Full audits retrieve every object; sampled and targeted audits retrieve $\min(100, n)$ and one object.}
\label{fig:audittime}
\end{figure}

\begin{figure}[t]
\centering
\includegraphics[width=\columnwidth]{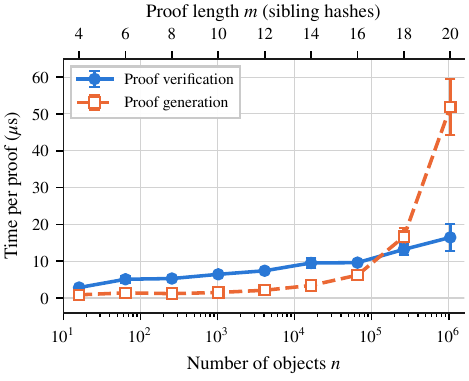}
\caption{Time per Merkle proof against the number of leaves (in-memory benchmark, 1{,}000 random proofs per run; mean $\pm$ s.d., 5 runs). Proof length is $\lceil\log_2 n\rceil$: 4 to 20 hashes, shown on the upper axis.}
\label{fig:verify}
\end{figure}

E5 repeats the commitment and the audits on the four datasets of Table~\ref{tab:datasets}; the component benchmark extends tree construction and proof operations to $2^{20}$ leaves. Table~\ref{tab:e5} and Figs.~\ref{fig:build}--\ref{fig:verify} show the results.

\begin{table*}[t]
\centering
\caption{E5: Scalability on the Object Store (Mean $\pm$ s.d.; 5 Runs, 3 for $n = 10{,}000$)}
\label{tab:e5}
\footnotesize
\begin{tabular}{@{}lrrrr@{}}
\toprule
 & $n = 10$ & $n = 100$ & $n = 1{,}000$ & $n = 10{,}000$ \\
\midrule
Hashing, all objects (ms) & 0.4 $\pm$ 0.0 & 4.0 $\pm$ 0.1 & 49.1 $\pm$ 1.0 & 455.1 $\pm$ 4.2 \\
Tree construction (ms) & 0.01 $\pm$ 0.00 & 0.09 $\pm$ 0.02 & 0.68 $\pm$ 0.17 & 6.73 $\pm$ 0.83 \\
Peak memory of commitment construction & 216 KiB & 431 KiB & 524 KiB & 1.70 MiB \\
Proof length, max / mean (hashes) & 4 / 3.60 & 7 / 6.88 & 10 / 9.98 & 14 / 13.63 \\
Proof generation, all $n$ (ms) & 0.05 $\pm$ 0.00 & 0.55 $\pm$ 0.02 & 7.0 $\pm$ 0.5 & 86.2 $\pm$ 2.4 \\
Proof verification, all $n$ (ms) & 0.05 $\pm$ 0.00 & 0.85 $\pm$ 0.08 & 11.4 $\pm$ 0.9 & 139.0 $\pm$ 1.5 \\
Merkle full audit (s) & 0.024 $\pm$ 0.002 & 0.145 $\pm$ 0.008 & 1.443 $\pm$ 0.153 & 14.293 $\pm$ 0.036 \\
Whole-root recomputation (s) & 0.015 $\pm$ 0.001 & 0.130 $\pm$ 0.011 & 1.441 $\pm$ 0.106 & 13.659 $\pm$ 0.249 \\
Full-data baseline (s) & 0.013 $\pm$ 0.002 & 0.136 $\pm$ 0.017 & 1.373 $\pm$ 0.088 & 13.526 $\pm$ 0.395 \\
Sampled audit, $c = \min(100, n)$ (ms) & 22.9 $\pm$ 2.0 & 138.3 $\pm$ 12.7 & 147.5 $\pm$ 13.4 & 153.3 $\pm$ 10.2 \\
Targeted single-object audit (ms) & 10.3 $\pm$ 1.9 & 9.8 $\pm$ 1.7 & 9.9 $\pm$ 2.0 & 10.6 $\pm$ 1.6 \\
\texttt{registerRoot} gas & 91,851 & 91,839 & 91,863 & 91,863 \\
\bottomrule
\end{tabular}
\end{table*}

\textit{Tree construction.} Construction time grows linearly with $n$: 0.68~ms for 1{,}000 and 6.73~ms for 10{,}000 objects, and in memory from 0.054~ms for 16 leaves to 722~ms for $2^{20}$ leaves (Fig.~\ref{fig:build}), approaching 0.7~$\mu$s per internal node for large trees. It is about 70 times cheaper than hashing the object contents (455~ms for 441.66~MiB), which is itself linear in the data volume. Memory is also linear: construction over $2^{20}$ leaves allocated 81.8~MiB at peak (internal nodes and level lists, excluding the leaves), because Python stores each 32-byte hash as a separate object.

\textit{Audit time.} Fig.~\ref{fig:audittime} separates the audit types. Full Merkle audits, whole-root recomputation and the full-data baseline all grow linearly and lie within about 6\% of one another (14.3, 13.7 and 13.5~s at 10{,}000 objects), because all three retrieve every object; the Merkle full audit is slightly slower, partly because it also generates and verifies $n$ proofs (1.6\% of its time at $n = 10{,}000$). A targeted audit stays at about 10~ms and a sampled audit of 100 objects at 138--153~ms from 100 to 10{,}000 objects, i.e. independent of $n$, because their work depends on $|C|$ and on $\log_2 n$ only. (At $n = 10$ and $n = 100$ the sample covers all objects and is a full audit.)

\textit{Proof generation and verification.} Proof length follows the bound of Eq.~\eqref{eq:prooflen}, from 4 to 20 hashes between 16 and $2^{20}$ leaves (Fig.~\ref{fig:verify}). Verification time follows it, from 2.9~$\mu$s to 16.5~$\mu$s per proof, i.e. about 0.8~$\mu$s per sibling. Proof generation is a walk of the same length, but its time rose from 0.9~$\mu$s to 52~$\mu$s, sharply above $2^{16}$ leaves. Because the number of lookups grows only from 4 to 20, we attribute the rise to memory access: a random path touches nodes spread over a structure of tens of megabytes. We did not isolate this effect.

\finding{5}{Tree construction and full audits are linear in $n$, and full audits cost the same as downloading the data. Proof size and verification grow logarithmically (20 hashes and 16.5~$\mu$s at $2^{20}$ objects), so targeted and sampled audits stay constant in time from 100 to 10{,}000 objects.}

\subsection{Communication Overhead (RQ4)}
\begin{figure}[t]
\centering
\includegraphics[width=\columnwidth]{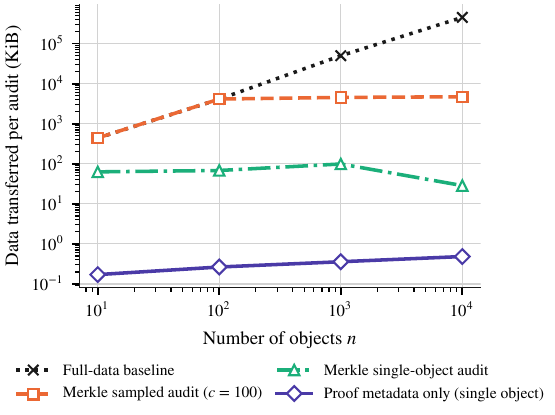}
\caption{Data transferred per audit against the number of objects (log--log). The single-object curve varies with the size of the randomly selected object; the proof alone grows from 175~B to 495~B.}
\label{fig:comm}
\end{figure}

Table~\ref{tab:comm} and Fig.~\ref{fig:comm} compare the bytes an auditor retrieves, measured as object content plus proof metadata.

\begin{table}[t]
\centering
\caption{Full-Data versus Merkle-Based Audit: Data per Audit}
\label{tab:comm}
\footnotesize
\setlength{\tabcolsep}{2.5pt}
\begin{adjustbox}{max width=\columnwidth}
\begin{tabular}{@{}lrrrr@{}}
\toprule
 & $n=10$ & $n=100$ & $n=1{,}000$ & $n=10{,}000$ \\
\midrule
Full-data baseline & 439.4 KiB & 4.00 MiB & 48.68 MiB & 441.66 MiB \\
Merkle full audit & 441.0 KiB & 4.03 MiB & 49.03 MiB & 446.27 MiB \\
\ \ proof share & 0.36\% & 0.64\% & 0.72\% & 1.04\% \\
Merkle sampled ($c=100$)$^\ast$ & 441.0 KiB & 4.03 MiB & 4.43 MiB & 4.60 MiB \\
Merkle single object & 62.8 KiB & 68.0 KiB & 98.8 KiB & 28.5 KiB \\
\ \ of which proof & 175 B & 271 B & 367 B & 495 B \\
Auditor-held state & 320 B & 3.1 KiB & 31.3 KiB & 312.5 KiB \\
\ \ (Merkle: none; root on-chain) & & & & \\
\bottomrule
\end{tabular}
\end{adjustbox}\\[2pt]
\raggedright\scriptsize $^\ast c=\min(100,n)$. Single-object values depend on the size of the randomly chosen object.
\end{table}

A Merkle full audit transfers slightly \emph{more} than the baseline, 0.36--1.04\% for keys and proofs, consistent with Eq.~\eqref{eq:bytes}. The reduction appears only when fewer objects are audited: at 10{,}000 objects a sampled audit of 100 objects transferred 4.60~MiB, 1.04\% of the dataset, and a targeted audit 28.5~KiB, of which the proof was 495~B (14 hashes plus bitmap and key). Moreover, the baseline auditor must hold 32 bytes per object (312.5~KiB at $n = 10{,}000$) and must protect that list, whereas the Merkle auditor holds nothing and reads 32 bytes of root from the contract.

\finding{6}{Merkle auditing does not reduce the data of a full audit; it adds about 1\%. It reduces transfer by letting an auditor verify one object (with a proof of 175--495~B) or a sample of objects against a constant-size public root, without holding any per-object state.}

\subsection{Smart Contract Evaluation (RQ5)}
\begin{figure}[t]
\centering
\includegraphics[width=\columnwidth]{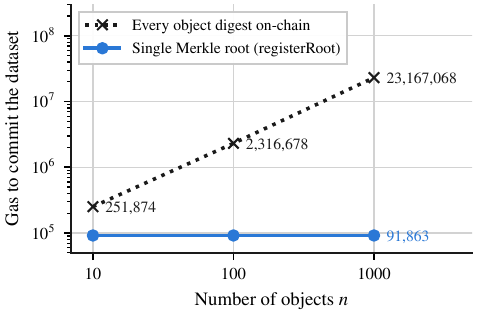}
\caption{Gas required to commit a dataset on-chain: one Merkle root versus all object digests (log--log). Point labels give the measured gas used (Table~\ref{tab:sc} and Section~\ref{sec:results}-G).}
\label{fig:gas}
\end{figure}

Table~\ref{tab:sc} reports gas and latency of the contract operations, and Fig.~\ref{fig:gas} compares committing a dataset by its root with storing every digest on-chain.

\begin{table}[t]
\centering
\caption{Smart Contract Evaluation (5 Runs)}
\label{tab:sc}
\footnotesize
\setlength{\tabcolsep}{3pt}
\begin{adjustbox}{max width=\columnwidth}
\begin{tabular}{@{}lrrr@{}}
\toprule
Operation & Gas used & Latency (ms) & Storage \\
\midrule
Deploy registry & 627,612 & -- & 2,656 B code \\
\texttt{registerRoot}, new dataset & 91,863 & 11.8 $\pm$ 0.8 & 3 slots \\
\texttt{registerRoot}, update & 37,885--37,897 & 12.2 $\pm$ 1.7 & -- \\
\texttt{recordAudit}, first & 118,104 & 14.4 $\pm$ 3.0 & 3 slots \\
\texttt{recordAudit}, subsequent & 101,016 & 11.7 $\pm$ 1.4 & 2 slots \\
\texttt{getRoot} (view)$^\dagger$ & 28,707 & 7.1 $\pm$ 0.8 & -- \\
\texttt{verifyProof} (view)$^\dagger$, 10 sib. & 40,473--40,517 & 7.3 & -- \\
\texttt{verifyProof} (view)$^\dagger$, 20 sib. & 55,633--55,681 & 9.9 & -- \\
\midrule
All digests on-chain, $n=1{,}000$ & 23,167,068 & 178.6 (10 tx) & 1,000 slots \\
\bottomrule
\end{tabular}
\end{adjustbox}\\[2pt]
\raggedright\scriptsize $^\dagger$\texttt{eth\_estimateGas}, includes 21,000 intrinsic gas; free when called off-chain. Latency on a local automining node.
\end{table}

Registering a dataset costs 91{,}863 gas whatever its size (91{,}839--91{,}863 for 10 to 10{,}000 objects; the small differences come from zero bytes in the calldata), almost all of it for writing three new storage slots; updating the root of an existing dataset costs 37{,}885--37{,}897 gas. Recording an audit costs 118{,}104 gas the first time (the record plus the array length) and 101{,}016 afterwards. Storing all 1{,}000 digests instead cost 23{,}167{,}068 gas in 10 transactions, 252 times the root, and grows linearly (251{,}874 gas for 10 and 2{,}316{,}678 for 100 objects). On-chain verification of a proof costs about 1{,}520 gas per sibling beyond a fixed part (26{,}852 for one sibling, 55{,}654 on average for 20), so even a dataset of $2^{20}$ objects can be verified inside a contract if needed; reading the root, which is all our off-chain auditor does, is free. Per audit, therefore, the blockchain adds one read (7--10~ms here) and optionally one transaction.

The latencies in Table~\ref{tab:sc} are those of a local node that mines every transaction at once. On a public network, a transaction is not included before the next block, and Ethereum mainnet divides time into 12-second slots~\cite{ethblocks}; gas is paid at a price set by the protocol's base fee and the sender's tip~\cite{eip1559}. Registration and audit records would then take seconds to minutes to confirm, while the root read stays a local RPC call.

\finding{7}{Keeping the trusted root on-chain costs 91{,}863 gas once per dataset version, independent of $n$, and a free read per audit; storing the digests themselves would cost 252 times more at 1{,}000 objects and grow linearly.}

\section{Security Analysis}\label{sec:security}

The integrity decision is always made the same way: current data $\rightarrow$ current digest and leaf $\rightarrow$ root reconstructed with the proof $\rightarrow$ comparison with the trusted root $\rightarrow$ decision. The blockchain does not detect tampering; it only guarantees that the value compared against has not changed since registration.

\textit{A. Modification.} If $D'_i \neq D_i$, then $H'_i \neq H_i$ except with probability negligible under the collision resistance of SHA-256, hence $\ell'_i \neq \ell_i$. To pass, the CSP would need sibling values that lead from $\ell'_i$ to $R^\ast$, which requires a collision in~\eqref{eq:node} somewhere on the path. Every modification in E2 and E3 was detected, consistent with this argument.

\textit{B. Multiple modifications.} Each object is verified through its own path against the same root, so several modifications neither cancel nor hide each other (E3: 2{,}205 of 2{,}205). With sampling, an unchallenged modification goes unnoticed in that audit; the residual risk is $1 - P_{\mathrm{det}}$ of Eq.~\eqref{eq:pdet}, and repeated audits with fresh random samples reduce it further. The challenge must be unpredictable to the CSP; our auditor draws it locally, and Zhang et al.~\cite{zhang2019procrastinating} show how block hashes can make it publicly verifiable.

\textit{C. Deletion.} The committed tree fixes the key and position of every object and the contract fixes $n$; a missing object therefore fails as \textsc{missing}, and removing or renaming its tree entry fails as \textsc{modified} (E4). Integrity auditing cannot restore the object; that requires replicas or backups.

\textit{D. Substitution, reordering and replay.} Because the leaf binds $\Hs(k_i)$, serving one object's content under another key, or swapping two objects, fails. Serving an older version of an object passes only if that version is the one committed by the current root; after a legitimate update and re-registration, stale content fails.

\textit{E. Integrity of the reference record.} Only the owner address can change a dataset's root, and every change increments the version and emits an event, so a CSP cannot replace the root and a legitimate change is visible. Audit records are append-only and include the root they were checked against.

\textit{F. Hash collisions and tree structure.} NIST assigns SHA-256 a collision strength of 128 bits~\cite{nist800107}; no practical collision is known for it, in contrast to SHA-1~\cite{stevens2017sha1}. Distinct prefixes for leaves and internal nodes~\cite{rfc6962} prevent presenting an internal node as a leaf, and promoting unpaired nodes avoids the ambiguity of trees that duplicate the last node, in which two different leaf sequences can share a root.

\textit{G. Smart-contract trust and root generation.} Blockchain immutability guarantees that the registered root is the one that was submitted, not that it was \emph{correct}. If the owner's environment or key is compromised, or the data was already corrupted when hashed, an attacker-chosen root is registered and the faulty data will pass every audit. Root generation must therefore run in a trusted environment over the intended data, and the owner key must be protected.

\textit{H. Limits of immutability and of the mechanism.} The contract cannot see off-chain data, so its guarantee is only as strong as the auditor's honest execution of Algorithm~\ref{alg:audit}, and it rests on the contract being free of the implementation flaws catalogued for Ethereum contracts~\cite{atzei2017survey,luu2016smarter}. The scheme provides neither availability nor confidentiality: a CSP that refuses to answer causes a failed audit but no recovery, and auditors read the plaintext of the objects they check; confidentiality would require encrypting objects before upload~\cite{kamara2010crypto}. A chain reorganization or a censored transaction can delay registration on a public network, which is outside our local evaluation.

\section{Discussion}\label{sec:discussion}

\textit{Answers to the research questions.} \textbf{RQ1:} unmodified objects always passed (E1), and single modifications of three kinds were detected and localized in 15 of 15 trials (E2). \textbf{RQ2:} full audits identified all 2{,}205 modified and all 830 deleted objects, separated the two classes and produced no false positives (E3, E4); sampled audits detect violations with the probability of Eq.~\eqref{eq:pdet}. \textbf{RQ3:} construction and full audits are linear in $n$, proof size and verification logarithmic, and targeted and sampled audits constant in $n$ (E5). \textbf{RQ4:} a full Merkle audit transfers about 1\% more than a full-data audit; a targeted audit needs one object and a proof of under 500~B, and a 100-object sample 1\% of the 10{,}000-object dataset. \textbf{RQ5:} the root costs 91{,}863 gas per dataset version independent of $n$, and an audit needs one free chain read.

\textit{Where the benefit lies.} The measurements correct a common expectation: a Merkle tree does not make checking \emph{every} object cheaper, because each object still has to be read and hashed, and in our testbed retrieval was 94\% of a full audit. Its benefits are elsewhere. It replaces a per-object reference list, which grows with the data and must itself be protected, by a 32-byte value that fits in a contract at constant cost (Fig.~\ref{fig:gas}); it lets anyone check one object, or a random sample, against that value at a cost independent of the dataset size; and it localizes each failure. Practical deployments would therefore combine frequent sampled audits, whose detection probability can be chosen through $c$, with occasional full audits.

\section{Threats to Validity}\label{sec:validity}

\textit{Internal validity.} All components share one 2-vCPU virtual machine, so object retrieval competes with the auditor and the chain node for CPU, and absolute retrieval times would differ with separate hosts. The S3 server needed a socket-option patch (Section~\ref{sec:impl}); without it, retrieval times were inflated by 40~ms per GET. Five repetitions (three at 10{,}000 objects) give modest statistical power; we report standard deviations, not confidence intervals. The prover ran in the auditor's process, so network transfer of proofs is represented by byte counts, not by measured time. Detection probabilities of sampled audits were estimated by drawing samples over the measured per-object outcomes, complemented by one real sampled audit per repetition.

\textit{External validity.} The object store was s3s-fs over loopback, not MinIO or a production cloud store; network latency, bandwidth limits, erasure coding and request pricing of real providers would raise retrieval time and cost, which strengthens rather than weakens the relative value of sampled audits. The blockchain was a local Hardhat network with instant mining; gas values are those of the EVM and carry over to Ethereum-compatible chains, but latencies do not. Datasets were synthetic, with object sizes from 1~KiB to 256~KiB and at most 10{,}000 stored objects (441.66~MiB); very large objects or millions of stored objects were evaluated only through the in-memory benchmark. Timings depend on the Python implementation.

\textit{Construct validity.} ``Detection time'' is the duration of a targeted audit of an object already known to be affected; time until a violation is noticed in practice depends on the audit schedule. Peak memory was measured with \texttt{tracemalloc} and covers Python allocations only.

\section{Limitations}\label{sec:limitations}
The evaluation is a local simulation, not a production cloud deployment. The security of the scheme depends on the root being generated over correct data and registered with an uncompromised key. Any change to the dataset requires rebuilding the affected path and registering a new root (37{,}885--37{,}897 gas per update here), so frequently changing data incurs on-chain cost per committed version; the prototype rebuilds the tree rather than updating it incrementally. Auditing detects but cannot restore deleted or modified objects, and it does not guarantee availability; recovery requires redundancy across providers, as in HAIL~\cite{bowers2009hail}. Auditors must download the objects they check, which reveals their contents and makes full audits as expensive as full downloads. On a public chain, transaction latency and gas fees would add seconds to minutes and monetary cost to every registration and recorded audit.

\section{Future Work}\label{sec:future}
We plan to repeat the evaluation against MinIO and a public cloud object store with the auditor on a separate host, and on an Ethereum test network to measure confirmation latency. The tree can be extended with incremental updates for insertions and deletions, and challenges can be derived from block hashes so that sampling is publicly verifiable. Combining the root with homomorphic tags~\cite{ateniese2007pdp,wang2011tpds} would allow auditing without downloading object content, and privacy-preserving auditing~\cite{wang2013tc} or zero-knowledge proofs could hide object contents from auditors. Incremental tree maintenance can draw on dynamic schemes~\cite{erway2009dpdp,yang2013dynamic}.

\section{Conclusion}\label{sec:conclusion}
This paper examined, through an implementation and repeated measurements, how far a deliberately simple design can go: SHA-256 leaves that bind each object's key to its content, one Merkle root per dataset, and a smart contract that keeps the root and an audit log. In the S3-compatible testbed with a local Ethereum network, no unmodified object was ever flagged, and every injected fault was found and attributed to the right object and class: 15 single modifications, 2{,}205 modifications in multi-object trials and 830 deletions. The costs separate cleanly. Building the tree and auditing every object grow linearly with the number of objects, and a complete audit is about as expensive as downloading the data. Checking a single object depends only on the logarithmic proof length (14 hashes, or 495~B including key and bitmap, at 10{,}000 objects), which kept targeted audits near 10~ms and 100-object samples near 150~ms at every dataset size we tried; the measured detection rates of sampled audits followed the hypergeometric model. Anchoring the root on-chain cost 91{,}863 gas per dataset version, 252 times less than writing 1{,}000 individual digests. The mechanism verifies integrity only; it neither keeps data confidential nor repairs it. All measurements come from a single-host simulation, and repeating them on production storage and a public blockchain remains future work.

\section*{Data Availability}
The source code, smart contract, unit tests, raw per-run JSON records of all experiments, the data series of every graph and the scripts that produce every figure and table in this paper are publicly available at \url{https://github.com/kgrcbit/Merkle-Tree-Based-Verifiable-Public-Data-Auditing}. The datasets are synthetic and are regenerated byte for byte from a fixed seed by the included experiment script.

\section*{Declaration of Competing Interest}
The authors declare that they have no known competing financial interests or personal relationships that could have appeared to influence the work reported in this paper.

\section*{Funding}
This research did not receive any specific grant from funding agencies in the public, commercial, or not-for-profit sectors.

\section*{Declaration of Generative AI and AI-Assisted Technologies}
During the preparation of this work the authors used Claude (Anthropic) to draft and revise the manuscript text, to implement the prototype and experiment scripts, and to execute the experiments reported in Section~\ref{sec:results}. After using this tool, the authors reviewed and edited the content as needed and take full responsibility for the content of the published article.

\end{document}